\documentclass[fleqn,usenatbib]{mnras}

\usepackage{newtxtext,newtxmath}

\usepackage[T1]{fontenc}

\DeclareRobustCommand{\VAN}[3]{#2}
\let\VANthebibliography\thebibliography
\def\thebibliography{\DeclareRobustCommand{\VAN}[3]{##3}\VANthebibliography}

\usepackage{graphicx}	
\usepackage{amsmath}	
\usepackage{anyfontsize}
\usepackage{adjustbox}
\usepackage{xcolor}
\usepackage[switch]{lineno}

\title[Chromatic Effects on PSF]{Chromatic Effects on the PSF and Shear Measurement for the \textit{Roman Space Telescope} High-Latitude Wide Area Survey }

\author[F. Berlfein et al.]{Federico Berlfein,$^{1}$\thanks{E-mail: fberlfei@andrew.cmu.edu}
Rachel Mandelbaum,$^{1}$
Xiangchong Li,$^{1,2}$
Tianqing Zhang,$^{3}$
Scott Dodelson,$^{1,4,5,6}$ and
\newauthor
Katarina Markovic$^{7}$
\\
$^{1}$ McWilliams Center for Cosmology and Astrophysics, Department of Physics, Carnegie Mellon University, Pittsburgh, PA 15213, USA \\
$^{2}$ Brookhaven National Laboratory, Bldg 510, Upton, New York 11973, USA\\
$^{3}$ Department of Physics and Astronomy and PITT PACC, University of Pittsburgh, Pittsburgh, PA 15260, USA
\\
$^{4}$ Department of Astronomy and Astrophysics, University of Chicago, Chicago, IL 60637, USA
\\
$^{5}$ Fermi National Accelerator Laboratory, P. O. Box 500, Batavia, IL 60510, USA
\\
$^{6}$ Kavli Institute for Cosmological Physics, University of Chicago, Chicago, IL 60637, USA
\\
$^{7}$ Jet Propulsion Laboratory, California Institute of Technology, 4800 Oak Grove Drive, Pasadena, CA 91109, USA
\\
}

\date{Accepted XXX. Received YYY; in original form ZZZ}

\pubyear{2015}

\begin{document}
\label{firstpage}
\pagerange{\pageref{firstpage}--\pageref{lastpage}}
\maketitle

\begin{abstract}
Weak gravitational lensing (WL) is a key cosmological probe that requires precise measurement of galaxy images to infer shape distortions, or shear, and constrain cosmology. Accurate estimation of the Point Spread Function (PSF) is crucial for shear measurement, but the wavelength dependence of the PSF introduces chromatic biases that can systematically impact shear inference. We focus on biases arising from spectral energy distribution (SED) differences between stars, used for PSF modeling, and galaxies, used for shear measurement. We investigate these effects in \textit{Roman's} four design reference mission WL bands (Y106, J129, H158, F184) and wide filter (W146). Using \textit{Roman}-like image simulations, we quantify the induced shear biases and compare them to requirements on those biases. Multiplicative biases over all galaxies hover around $\sim$0.2\% in the WL bands and 2\% in the wide filter, exceeding the mission requirement of $|m| < 0.032\%$ and relaxed requirement of $|m| < 0.1\%$. In individual redshift bins, biases can reach 0.4–0.9\% for the WL bands and 3–6\% for the wide filter. Additive biases remain acceptable in the WL bands but exceed systematic limits in the wide filter. We develop and test PSF-level corrections, showing that a first-order correction reduces biases within survey requirements for the WL bands; however, higher-order terms are necessary for the wide filter. Our results highlight the necessity of chromatic corrections for precision WL with \textit{Roman} and provide a framework for mitigating these biases. Finally, we compare analytical color-based corrections to self-organizing maps (SOMs) and find that both methods effectively reduce biases. 
\end{abstract}







\section{Introduction}

Weak gravitational lensing (WL) manifests as the correlated distortions, or shear, in the
shapes of distant galaxies caused by the gravitational field of intervening foreground matter \citep{Hoekstra_2008, Bartelmann_2017, Mandelbaum_2018}. The deflection of light by massive structures distorts the observed shapes of distant galaxies, allowing us to infer statistical properties of the underlying mass distribution. This makes WL sensitive to the expansion and growth history of the Universe, making it a crucial probe for cosmology and large-scale structure. The Dark Energy Survey \citep[DES;][]{DES_2005}, Hyper Suprime-Cam Survey \citep[HSC;][]{Aihara_2017}, and Kilo-Degree Survey \cite[KiDS;][]{de_Jong_2012} have made the most recent and precise measurements of cosmic shear, constraining the amplitude of the matter fluctuations to the percent-level \citep[see][]{Asgari_2021,Amon_2022, Secco_2022, Dalal_2023, Li_HSC}. However, upcoming surveys like the Vera C.\ Rubin Observatory Legacy Survey of Space and Time \citep[LSST;][]{LSST_2019}, \textit{Euclid} \citep{Euclid}, and the \textit{Nancy Grace Roman Space Telescope} \citep{Spergel_2015} will provide an unprecedented increase in survey size for WL, enabling greater statistical precision and cosmological constraining power. 

The \textit{Roman} High-Latitude Wide Area Survey (HLWAS) presents a unique opportunity for WL studies due to its unprecedented combination of area, depth, and resolution in the near infrared (NIR). Four NIR bands are expected to be used for WL according to the design reference mission (DRM)\footnote{The DRM can be found here: \url{https://roman.gsfc.nasa.gov/science/workshop112021/presentations/Mon_Overview/roman_mission.pdf}. The Roman Observations Time Allocation Committee (ROTAC) final report and recommendations came out on April 24th, 2025 and proposed multiple changes to the HLWAS described in the DRM, including the removal of imaging in F184 from the area used for WL. Given that this report was submitted after the completion of this work, we leave the implementation of different filter and survey configurations to future work. The ROTAC report can be found here: \url{https://roman.gsfc.nasa.gov/science/ccs/ROTAC-Report-20250424-v1.pdf}.}: Y106, J129, H158, and F184, covering a broad wavelength range between 0.9-2$\mu$m. It is important to note that the DRM does not represent the final survey strategy for \textit{Roman}. Compared to ground-based telescopes, \textit{Roman} benefits from the absence of atmospheric effects, allowing for significantly improved stability and modeling of optical effects \citep{Liaudat_2023}. This makes \textit{Roman} particularly well-suited for precision WL. Moreover, the high-resolution imaging provided by \textit{Roman} enables the recognition of blended galaxies that are  difficult to distinguish with ground-based telescopes \citep{Troxel_2022}. \textit{Roman’s} observations will also complement those from the LSST, enabling synergies between space- and ground-based WL analyses \citep{Eifler_2021}. While LSST provides deep, multi-band optical imaging over a much wider field, \textit{Roman}'s high-resolution imaging and coverage of the NIR will allow for improved detection, redshift estimation, and calibration \citep{Eifler_2021}.

Nevertheless, the increase in statistical precision means calibration of systematic effects will become extremely important \citep{Albrecht_2006, Schaan_2017}. Accurate WL analysis requires precise measurement of galaxy shapes, which in turn depends on the accurate characterization and modeling of the Point Spread Function (PSF). The PSF describes the response of the system to a point source. This includes optical elements, detector effects, pixelization, and atmospheric effects in the case of ground-based telescopes. Any misestimation of the PSF can introduce systematic biases that propagate into shape measurement, ultimately biasing cosmic shear and cosmological analysis \citep{Mandelbaum_2018}. The PSF is position- and wavelength-dependent, and proper modeling for both dependencies is needed for accurate PSF modeling \citep[e.g.,][]{Schutt_2025}. The wavelength dependence of the PSF for diffraction-limited surveys like \textit{Roman} is much stronger than that of atmospheric-dominated surveys \citep{Cypriano_2010}, posing challenges and questions about possible chromatic effects on the PSF.


 Several studies have investigated chromatic PSF biases in weak lensing, focusing on different sources and mitigation strategies. \cite{Cypriano_2010} and \cite{Eriksen_2018} demonstrated that differences between the spectral energy distributions (SEDs) of stars and galaxies introduce significant shear biases. Since the PSF is the response of the imaging system to a point source, and galaxies are extended objects, the PSF can be measured only using stars. However, galaxy shape measurement requires the use of the galaxy PSF, not the stellar PSF. This poses a problem: since stars and galaxies have systematically different SEDs, using stars to model the PSF means that the model does not describe the PSF for galaxies. 
 This in turn introduces a chromatic bias to shear measurement. We propose and validate methods such as color matching and template fitting to mitigate these effects. 
 As an additional complication, galaxy color gradients mean that there is actually not a single well-defined PSF that convolves galaxy light profiles.  \cite{Voigt_2012} and \cite{Semboloni_2013} examined galaxy color gradients, showing that internal SED variations contribute to biases, which can be reduced using narrow filters or multi-filter imaging. Atmospheric effects, including differential chromatic refraction and chromatic seeing, were explored by \cite{Meyers_2015a} and \cite{Plazas_2012}, who proposed PSF-level corrections to account for these biases in ground-based surveys like LSST. Finally, \cite{Meyers_2015b} studied the wavelength dependence of CCD sensitivity, highlighting the importance of modeling detector effects to avoid biases in shear measurements. Together, these works underscore the necessity of precise chromatic corrections for weak lensing surveys.

In this work, we focus primarily on the problem of star-galaxy SED differences on PSF estimation and shear measurement for \textit{Roman}. Moreover, due to the nature of our simulations, we implicitly test the impact of galaxy color gradients as well.  
We will require the use of realistic stellar and extragalactic catalogs with simulated SEDs. For this we use the stellar and extragalactic catalog used for the recently released Rubin-\textit{Roman} joint simulations, OpenUniverse2024 \citep{OU_2025}. Given the sensitivity of our study to the SED library, we also used the cosmoDC2 extragalactic catalog for comparison. To test the impact of these effects, we produce \textit{Roman}-like image simulations consisting of individual galaxy and star postage stamps with realistically complex SEDs and chromatic PSFs.  
These simulations use up-to-date models for the \textit{Roman} PSF, which should be reasonably close to the PSF in real \textit{Roman} data.

We will first quantify, and then contextualize within WL requirements, the magnitude of multiplicative and additive shear biases introduced by chromatic effects. This will serve the initial purpose of understanding the size of the effect and how it compares to the requirements for \textit{Roman}. This will be done for the 4 WL bands from the DRM and the wide filter (W146). The use of the wide filter, covering the same wavelength range as the 4 WL bands, would increase the depth and signal-to-noise ratio (SNR), allowing much fainter galaxies to be observed with higher precision for fixed exposure time. This would increase the effective number of galaxies used for WL, providing better statistics. However, it comes at the cost of increased systematics, since the wide filter would be more sensitive to chromatic effects. Photometric redshift estimation would also be impacted in a survey utilizing a single wide filter due to the absence of complementary NIR color information. We will quantify the magnitude of these effects for the wide filter as input to consideration of its potential use for WL. We will show that these biases are significant for the \textit{Roman} WL bands and require calibration/correction. We then develop and test mitigation methods that leverage our knowledge of the \textit{Roman} PSF to reduce chromatic biases to within the survey's systematic error budget. Finally, we will compare mitigation performance under semi-realistic conditions to better understand the potential accuracy of the methods developed.

The remainder of this paper is structured as follows. In Sec.~\ref{Background}, we provide background information on weak lensing and PSF modeling, emphasizing the importance of chromatic effects. Sec.~\ref{Simulations} describes the extragalactic and stellar catalogs used, the noise generation and image simulation process, and the software used for shear measurement. Sec.~\ref{Bias} presents the results of our analysis, quantifying shear biases induced by chromatic effects in different \textit{Roman} filters and assessing their impact on weak lensing science. Sec.~\ref{Mitigation} introduces and validates the proposed PSF-level mitigation method under perfect conditions. In Sec.~\ref{Results}, we introduce and compare different mitigation techniques, including color-based analytical corrections and machine-learning-based methods. Finally, in Sec.~\ref{Conclusion}, we summarize our findings, discuss the broader implications of our results, and outline potential future directions for improving chromatic bias corrections in weak lensing surveys.

\section{Background}\label{Background}

\subsection{Weak Lensing and Shear Bias}\label{background:Wl and shear bias}


Cosmological weak lensing (WL) is a powerful tool that relies on the measurement of distortions in the shapes of distant galaxies \citep{Hoekstra_2008, Bartelmann_2017, Mandelbaum_2018}. These distortions, quantified as the shear $\gamma$, are a primary observable in WL studies\footnote{Shear here refers to shape distortions caused by gravitational lensing only. However, intrinsic alignments \citep[IA; see ][]{Joachimi_2015, Troxel_2015}—the coherent alignment of galaxy shapes either with each other or with the surrounding density field—violate this assumption and therefore contribute to the weak lensing signal. In this work, we do not consider effects from IA, but rather it will be modeled as a contribution to the signal as in current cosmological lensing measurements.}. Caused by the gravitational potential of intervening matter along the line of sight, they allow us to probe the structure and matter content in the Universe.

To first order, the observed ellipticity, $\epsilon$, of a galaxy is related to the reduced shear, $g$, and the intrinsic shape of the galaxy, $\epsilon_s$, by:
\begin{equation}\label{eq:obs_e}
\epsilon \approx \epsilon_s + g,
\end{equation}
where $g = \gamma / (1 - \kappa)$ and $\kappa$ is the convergence \citep{Bartelmann_2017}. 
The convergence, $\kappa$, represents the integrated overdensity along the line of sight and is directly related to the gravitational potential of the intervening mass \citep{Munshi_2008, Mandelbaum_2018}. This relation connects the distortion of galaxy shapes to the underlying distribution of matter in the Universe. In the limit of WL it is safe to assume $|\kappa| \ll 1 $, meaning we can take $g \approx \gamma$ \citep{Hoekstra_2008}. Note that shear ($\gamma$) and reduced shear ($g$) are complex quantities with two components each (e.g., $\gamma = \gamma_1 + i\gamma_2$), corresponding to distortions along different axes. In that case, assuming galaxies have random intrinsic orientations gives a linear relationship 
between the average ellipticity, $\langle \epsilon \rangle$, and $\gamma$.

In the context of modern cosmological analysis, two-point correlation functions of the shear and galaxy density field, such as the shear auto-correlation function $\xi_{\pm}$ (cosmic shear), density autocorrelation (galaxy clustering) and their cross-correlation (galaxy-galaxy lensing), are key observables used to probe the underlying matter density field and constrain cosmology \citep{Hoekstra_2008, Kilbinger_2015}. Measuring shear accurately is therefore crucial for reliable cosmological measurements. Observationally, the measured shear, $\hat{\gamma}$, can be biased by systematic effects such as PSF and galaxy shape measurement errors, blending and detection bias \citep{Massey_2012, Mandelbaum_2018}.  
It is common to express the relationship between the observed and true shear through a linear model \citep{Huterer_2006, Heymans_2006}:
\begin{equation}\label{eq:obs_shear}
\hat{\gamma} = (1 + m) \gamma + c,
\end{equation}
where $m$ and $c$ represent multiplicative and additive biases, respectively. These biases can be introduced separately by different effects. For example, misestimations of the PSF size are often associated with multiplicative biases, 
while additive bias can stem from residual anisotropies in the PSF \citep{Massey_2012}. 
In terms of cosmological impact, a multiplicative bias of 1 per cent ($|m| = 0.01$) roughly translates to a 1.5 per cent bias in $S_8 = \sigma_8 \sqrt{\Omega_m/0.3}$ \citep{Yamamoto_2022}, where $\sigma_8$ quantifies the amplitude of matter fluctuations and $\Omega_m$ represents the fractional energy density of all matter. This gives a direct way of interpreting the biases in galaxy shape measurements with cosmological biases. Therefore, understanding potential sources of bias in shear measurements is crucial in obtaining unbiased cosmological measurements.

\subsection{PSF modeling}


Accurate PSF modeling \citep{Anderson_2000, Piotrowski_2013,Jarvis_2021, Liaudat_2023, Schutt_2025} is essential in WL studies because the PSF affects the observed shapes of galaxies, and therefore the inferred shear. Factors influencing the PSF include optical aberrations \citep{Wyant_1992}, detector effects \citep[e.g.,][]{Antilogus_2014,Plazas_2016}, and for ground-based telescopes, atmospheric effects \citep{deVries_2007, Chang_2012}. These effects combine to produce a position- and wavelength-dependent PSF, $\text{PSF} (x,y,\lambda)$. For space-based telescopes like \textit{Roman}, the primary source of chromaticity in the PSF comes from diffraction due to the finite aperture of the telescope, filter coatings and substrate refraction. This results in a roughly proportional dependence between PSF size and wavelength \citep{Cypriano_2010,Kamath_2019}.  
In contrast, the chromaticity for ground-based telescopes, like LSST, is dominated by Kolmogorov turbulence in the atmosphere and results in a much weaker wavelength dependence of $\sim \lambda^{-0.2}$ for the PSF size \citep{Roddier_1981, Meyers_2015b}. In addition, the wavelength-dependent transmission for the Roman filters is being characterized with laboratory tests, accounting for angle of incidence effects and variations in filter coating thickness (Switzer et al., in prep).
For these reasons, chromatic effects in \textit{Roman} are especially important for accurate PSF modeling.


The lack of atmospheric effects for space-based telescopes allows for very accurate modeling of the physical effects that go into the \textit{Roman} PSF (optical and detector effects). Ray-tracing simulations and detector tests have enabled WebbPSF \citep{Perrin_2014} 
and the \texttt{galsim.roman} \citep{Kannawadi_2016} module
to provide high-fidelity models of the \textit{Roman} PSF already. These models allow us to create realistic image simulations \citep{Troxel_2022, OU_2025} for analysis pipeline testing and studies of systematic biases and uncertainties. However, empirical modeling of the PSF through real observations is still needed for shape measurement. This process involves using star images to model the PSF across the focal plane. However, what we observe in real images is not the wavelength-dependent PSF, $\text{PSF} (x,y,\lambda)$, but rather the effective PSF:
\begin{equation}\label{eq:eff_psf}
    \text{PSF}_{\text{eff}, o} (x,y) = \frac{\int \mathrm{d}\lambda \;  \text{PSF} (x,y,\lambda)\; F(\lambda) \; \text{SED}_o(\lambda)}{\int \mathrm{d}\lambda \; F(\lambda) \; \text{SED}_o(\lambda)},
\end{equation}
where $F(\lambda)$ is the filter throughput and $\text{SED}_o(\lambda)$ is the spectral energy distribution of some object $o$. Conventionally, the units for the SED are in erg/cm$^2$/s/\AA. However, one needs an additional factor of $\lambda/hc$ inside of the integrals in order to convert ergs to photons, which is what actually gets counted in the detector. Note that we have implicitly included the effects of the pixel response function (i.e., the convolution with a top-hat filter corresponding to the pixel size) in $\text{PSF} (x,y,\lambda)$. 
The normalization factor, given by the flux, assures that the integral over the image space sums to 1 (assuming the image extends infinitely).

We need the effective PSF to measure galaxy shapes, but since galaxies are not in the same positions as stars and do not share the same SED, both spatial and chromatic interpolation of the effective PSF is needed for galaxy shape measurement. This poses a significant challenge in PSF modeling as both spatial and chromatic effects can be correlated. However, for \textit{Roman} we assume that these effects can be treated independently and the spatial and chromatic interpolation can be done separately. Therefore, in this work we focus on understanding the impact of PSF errors caused by SED differences between the stars used for modeling and the galaxies used for measurement. We will work under the assumption of perfect spatial interpolation to capture only chromatic differences. The chromatic PSF errors can then be expressed as:
\begin{align}\label{eq:eff_psf_diff}
    \Delta \text{PSF}_{\text{eff}} (x,y) &=  \text{PSF}_{\text{eff}, \star} (x,y) -\text{PSF}_{\text{eff}, \text{g}} (x,y)\\
    &= \int \mathrm{d}\lambda \;  \text{PSF} (x,y,\lambda)\; F(\lambda) \; \left[S_{\star}(\lambda) - S_{\text{g}}(\lambda)\right]\nonumber
\end{align}
where the $\star$ and ``g'' subscripts represent the stellar and galaxy quantities, respectively, and $S_{\star}$ and $S_{\text{g}}$ are the flux-normalized SEDs. In practice the stellar PSF is constructed from a collection of stars, so $\star$ really corresponds to an ensemble of stars rather than any individual star.

\subsection{Previous Work on Chromatic Biases}

Chromatic effects arising from the wavelength dependence of the PSF can represent a significant source of systematic error in weak lensing measurements. These biases can originate from multiple sources, including the intrinsic properties of galaxies (e.g., SED differences), atmospheric effects (e.g., differential chromatic refraction and chromatic seeing), and detector effects (e.g., CCD response). Several methods have been proposed to model and mitigate these biases. In this section, we review key contributions that address chromatic biases in weak lensing surveys, highlighting the challenges posed by different sources of chromaticity and the effectiveness of proposed mitigation strategies.

\subsubsection{SED Effects}

\cite{Cypriano_2010} and \cite{Eriksen_2018} investigated the impact of the wavelength dependence of the PSF for diffraction-limited instruments like \textit{Euclid}. Both studies demonstrated that differences in the SEDs of stars and galaxies can lead to significant shear biases if the PSF is not properly modeled. Two mitigation methods were evaluated in \cite{Cypriano_2010}: broad-band color matching and template fitting. Color matching uses stars with colors similar to those of galaxies to estimate the PSF. While this approach reduces biases, it is limited by the intrinsic differences in the PSF size-color relation between stars and galaxies and struggles at high redshift. Template fitting predicts the PSF size for each galaxy based on its estimated SED from multi-band photometry. This was shown to be more robust, particularly at high redshifts, and reduces residual biases more effectively.


\cite{Eriksen_2018} evaluated both template-fitting and machine-learning approaches to estimate the effective PSF size, analyzing their sensitivity to photometric calibration errors, wavelength-dependent PSF models, and SED template libraries. They demonstrated that conventional template-fitting methods, which rely on photometric redshifts and galaxy SED models, are prone to biases from photometric uncertainties and limited template coverage. These biases are particularly pronounced for galaxies with poorly constrained redshifts or SEDs. To address these limitations, the authors proposed a hybrid approach combining machine learning and calibration data from archival observations, achieving substantial reductions in PSF size biases. Their results highlighted the necessity of accounting for correlations between photometric redshifts and PSF estimates, as neglecting these can exacerbate systematic biases.

\cite{Voigt_2012} and \cite{Semboloni_2013} studied the impact of galaxy color gradients, a second-order chromatic effect that arises due to spatial variations in the galaxy SED. They simulated galaxy images with bulge and disk components characterized by distinct SEDs and evaluated shear biases introduced when PSF corrections are based solely on the composite galaxy spectrum. Their findings showed that the magnitude of the bias is strongly dependent on galaxy and survey properties such as the bulge-to-disk size ratio, the internal color gradient, PSF-to-galaxy size ratio and filter width. However, both studies acknowledged that for a `typical' galaxy, expected \textit{Euclid} PSF FWHM and filter width, the expected bias is subdominant to other effects and within survey requirements ($|m| \sim 10^{-4}$).


\subsubsection{Atmospheric Effects}

\cite{Meyers_2015a} explored the impact of atmospheric chromatic effects, specifically differential chromatic refraction (DCR) and the wavelength dependence of seeing, for LSST. They highlighted that these atmospheric chromatic effects result in biases in galaxy shape measurements due to the SED differences between stars and galaxies. They demonstrated that DCR shifts the centroid of the PSF depending on the zenith angle and the SED, while chromatic seeing alters the PSF size and ellipticity. The resulting biases in inferred shear, if uncorrected, can exceed LSST's systematic error requirements. To mitigate these effects, the authors proposed a PSF-level correction method that involves applying perturbations to the stellar PSF to model the galaxy PSF. DCR corrections involve deconvolving the stellar PSF in the zenith direction with a Gaussian kernel, whose second moment depends on the SED differences between the observed star and a fiducial monochromatic SED. Chromatic seeing corrections are applied by scaling the PSF's coordinate axes based on the ratio of PSF sizes predicted from photometric data using an \texttt{Extra Trees Regression} model. The corrected fiducial PSF is then interpolated to galaxy positions, and the process is reversed for galaxies to construct their effective PSFs. The authors then used the formalism in \cite{Paulin_2008} to relate  the PSF size and shape errors to shear multiplicative and additive biases, respectively. Tests on image simulations showed that residual biases in shear measurements are reduced to levels well within LSST's systematic error budget.

\cite{Plazas_2012} investigated the impact of atmospheric dispersion, caused by the wavelength-dependent refraction of light, on shear measurements for DES and LSST. Atmospheric dispersion elongates images along the zenith direction, introducing biases in galaxy shapes that mimic cosmic shear signals if not properly corrected. They find that these biases exceed the statistical error budgets for DES in the \(g\) and \(r\) bands, and for LSST in the \(g\), \(r\), and \(i\) bands, with the largest errors occurring in bluer bands. To mitigate these effects, they propose a linear correction to the PSF size and centroid based on galaxy color, calibrated empirically using stars or theoretically derived for galaxies (dispersion for galaxies can't be measured empirically). This correction significantly reduces systematic errors in the \(r\) band for DES and the \(i\) band for LSST, though residual biases remain significant for LSST's \(r\) band. For the \(g\) band, atmospheric dispersion effects dominate, suggesting that cosmic shear measurements should rely on redder bands for accuracy.

\cite{Carlsten_2018} investigated the wavelength dependent PSF in the Hyper Suprime-Cam (HSC) Subaru Strategic Program (SSP) survey. Using HSC data, they measured the PSF size as a function of stellar color across multiple bands ($g$, $r$, $i$, $z$, and $y$) and found that redder stars exhibited smaller PSF sizes than bluer stars, with the effect being strongest (1-2\%) in the $g$, $r$, and $i$ bands, and negligible in the $z$ and $y$ bands. These trends are consistent with atmospheric turbulence and partially influenced by instrumental contributions such as charge diffusion in CCDs. To address these chromatic biases, they model the PSF as a power-law function of wavelength, \(\text{PSF}(\lambda) \propto \lambda^{-b}\), and fit the parameter \(b\) using stellar data. They find that \(b\) varies with observing conditions, field position, and filter, requiring accurate calibration to mitigate biases. Simulations demonstrate that with sufficient stellar density (\( \sim 1 \, \text{star arcmin}^{-2} \)), the parameter \(b\) can be constrained to \(\Delta b \approx 0.02\) for upcoming surveys like LSST, which is within the survey's systematic error budget. Similarly, \cite{Zhang_2018} measures the parameter $b$ using data from the Parallel
Imager for Southern Cosmological Observation (PISCO) and finds $b\sim0.25 \pm 0.03$.

Although space-based telescopes like \textit{Roman} do not experience atmospheric effects, the resulting impacts on the PSF and how to mitigate them can translate to the sources of systematic bias we care about in space-based telescopes. For example, \cite{Meyers_2015a} show how chromatic seeing alters the PSF size and mitigate the effect by using photometric data and ML techniques to perform a correction. Chromatic effects that alter the PSF size can be present in space-based telescopes, and so understanding how they can be corrected, regardless of the source of the bias, can be extremely helpful in developing new techniques.

\subsubsection{Detector Effects}

\cite{Meyers_2015b} investigated chromatic effects caused by the wavelength dependence of CCD absorption depths in LSST. The wavelength-dependent absorption depth of photons in silicon leads to chromatic changes in the PSF size and ellipticity, particularly in the redder bands. The authors found that while subdominant to atmospheric contributions, this effect can still bias shear measurements significantly within the LSST's systematic error budget if left uncorrected. To mitigate these effects, the authors proposed that one can compute and remove chromatic PSF misestimation on a galaxy-by-galaxy basis, given proper knowledge of the SEDs and wavelength dependence of the PSF. They concluded that with proper modeling of CCD chromatic effects and leveraging of LSST's six-band photometry, biases can be reduced to levels below the survey requirements.

\textit{Roman} introduces new challenges as it uses H4RG-10 detectors rather than CCDs. Tremendous efforts have already been made to characterize the properties of the \textit{Roman} detectors \citep[e.g.,][]{Mosby_2020}. Work is ongoing to understand the instrumental chromatic response, including detector effects, filter transmission, and telescope optics (Switzer et al., in prep). For example, we anticipate wavelength-dependent charge diffusion effects that have been measured in \textit{Euclid} CCDs \citep{Niemi_2015} may also be important for \textit{Roman}’s H4RG detectors.

\subsubsection{Unexplored Avenues}
Significant progress has been made in understanding and modeling the effects of chromaticity in the PSF for surveys like LSST and \textit{Euclid}. However, the impact of these effects on the \textit{Roman} telescope remains largely unexplored. 
In this work, we specifically focus on how shear measurement in Roman is impacted when the PSF is modeled using stars (with stellar SEDs) but used to measure galaxy shapes (with galaxy SEDs). 
This type of analysis has yet to be conducted for \textit{Roman} and is expected to be a non-negligible effect given the stringent shear calibration requirements of Stage IV surveys\footnote{Stage IV surveys should collectively achieve or exceed a factor of 10 gain over Stage II experiments in the dark energy figure of merit \citep{Albrecht_2006}.}. In addition, given the intrinsic SED differences when looking at NIR or visible spectrum, it is important to have an analysis specific to the \textit{Roman} bands. Quantifying this effect and demonstrating effective mitigation schemes is of paramount importance and is the focus of this work. In addition, due to how our simulations are constructed, we implicitly test the impact of galaxy color gradients on shape measurement in the context of chromatic PSFs. This builds on previous work in this area, but for the case of \textit{Roman} rather than \textit{Euclid}.


\section{Simulations and Shear Estimation}\label{Simulations}

This section details the data and software used in this work. We start by describing the input extragalactic and stellar catalogs.  We then discuss the software used to generate semi-realistic noisy magnitudes at the catalog-level, along with the selection cuts applied using these ``observed'' quantities. Finally, we describe the software tools and choices that went into the image simulation and shear measurement process.

\subsection{Extragalactic Catalogs}
In this subsection we discuss the two extragalactic catalogs that we used: cosmoDC2 \citep{Korytov_2019} and \texttt{Diffsky} \citep{OU_2025}, chosen due to their distinct and realistically complex SED libraries. Since chromatic effects are intrinsically dependent on the SED, using both catalogs allows us to investigate the impact of the choice of SED library. Both catalogs are divided and stored in equal-area tiles using \texttt{HEALPY} \citep{Healpy}, with NSIDE = 32. To reduce computational expense given the size of the full catalogs, only a single tile from each catalog is chosen for galaxy selection. A single tile covers approximately 3 square degrees, providing a sufficiently large area to avoid unintentionally selecting a galaxy cluster or void region.

\subsubsection{CosmoDC2}
The cosmoDC2 \citep{Korytov_2019} extragalactic catalog \citep{Kovacs_2022} was developed using a data-driven approach to semi-analytical modeling of the galaxy population, as applied to the large-volume Outer Rim simulation \citep{Heitmann2019}. The catalog combines resampled results from the Galacticus semi-analytical model \citep{Benson2012} with empirical methods to ensure consistency with observational data, enabling realistic galaxy property variations with redshift and appropriate levels of galaxy blending for cosmological analyses. Galaxies are modeled as a combination of bulge and disc components, each with their own SED and described by Sersic profiles with parameters $n = 4$ and $n=1$, respectively. To capture the complexity of star-forming regions in galaxy discs, a random walk component was introduced, redistributing some disc flux into point sources with the same SED, spatially distributed according to a Gaussian profile matching the disc's size and shape. In this work we ignore the knots and preserve the original disc flux, modeling the galaxy only as a bulge + disk.

The SED library in cosmoDC2 includes four distinct types: Burst, Instantaneous, Constant, and Exponential -- designed to model a wide range of galaxy star formation histories.
Burst SEDs represent galaxies that experienced short, intense star formation events. Instantaneous SEDs assume all stars formed at a single time, modeling the evolution of a stellar population after an isolated formation episode. Constant SEDs simulate galaxies with continuous star formation at a steady rate over time. Exponential SEDs describe galaxies in which star formation rates decline exponentially with time. Each SED type contains templates parameterized by galaxy age and stellar metallicity, allowing for realistic representation of galaxies at different evolutionary stages. 
These templates are assigned to galaxies based on their physical properties, ensuring that the simulated photometry accurately reflects observed galaxy populations.

For this work, Healpix tile \texttt{10067} was chosen at random. This tile contains a total of 7,210,641 galaxies, of which we select a random subset of 1 million galaxies for our simulations. From this subsample, there are a total of 637 unique SED templates used, drawn in roughly equal numbers from the four SED types.


\subsubsection{Diffsky}

The \texttt{Diffsky} \citep{OU_2025} extragalactic catalog was constructed using the GalSampler technique \citep{Hearin2020}, which transfers synthetic galaxy populations from the Outer Rim \citep{Heitmann2019} high-resolution $N$-body simulation to a larger-volume simulation via a halo-to-halo correspondence. This process includes populating host halos with central and satellite galaxies, followed by enriching the galaxy population with additional properties. While this methodology is similar to that used for the cosmoDC2 catalog \citep{Korytov_2019}, the new catalog incorporates updated models for assigning galaxy properties, many of which are prototypes from a differentiable and probabilistic model of the galaxy–halo connection\footnote{\label{lsst_diffsky}\url{https://diffsky.readthedocs.io/en/latest/}}. 
As for cosmoDC2, galaxies in \texttt{Diffsky} are modeled as a bulge, disk, and knots, but we ignore the knots and redistribute that flux back into the disk component.

The SEDs and photometry of galaxies in the catalog are computed using stellar population synthesis (SPS) models with the DSPS library \citep{Hearin_2023}, incorporating models for stellar metallicity, dust attenuation, and burstiness. Galaxy SEDs are derived by convolving simple stellar population (SSP) templates from the MILES library \citep{FalconBarroso2011} with probability distributions of stellar age and metallicity, while dust attenuation effects are modeled using a parametric approach that varies with stellar mass, star formation rate, and stellar population age. Additionally, short-timescale star formation burstiness is captured through a parametric model that describes recent star formation contributions using scaling relations dependent on stellar mass and specific star formation rate. This process differs significantly from that used in cosmoDC2, resulting in a distinct SED library with more realistic colors across optical and NIR wavelengths \citep{OU_2025}. Additionally, each galaxy has its own unique 
SED (for each component), meaning we cannot directly associate it with a specific SED template from a fixed library.
Further details on SEDs and \texttt{Diffsky} can be found in \cite{OU_2025} and the publicly released \texttt{lsstdesc-diffsky}\footnote{See footnote \ref{lsst_diffsky}} code.

We chose to use the first publicly available data release of this catalog, the single Healpix tile \texttt{10307}.

\subsection{Stellar Catalog}

We use the stellar catalog from the most recent Rubin-Roman simulations \citep{OU_2025}. The stellar population comes from Galfast \citep{Juri_2008} and is based on an extrapolation of the observed SDSS stellar catalogs. The SEDs are matched to the predicted colors using different stellar models. Kurucz models \citep{Kurucz1993} are used for main sequence stars and giants, \cite{Bergeron1992} models for white dwarfs, and a combination of spectral templates and SDSS observations for M, L, and T dwarfs. We choose only stars with SEDs from the Kurucz library, as these represent the stellar types that are used for PSF modeling. Given this downsampling choice, we use the entire simulation footprint (i.e., from all tiles) in order to obtain 400,000 unique stars, which is the number needed for our simulations (see Sec.~\ref{subsec:sims}). There are a total of 4885 unique Kurucz SED templates from this library.  For normalization, every star in the catalog includes a reference normalization magnitude at 500~nm.


\begin{table}
\begin{adjustbox}{width=0.8\columnwidth,center}
\begin{tabular}{ c|c|c| }

\multicolumn{3}{c}{Filter-dependent Error Model Parameters} \\
\hline
Filter &  5$\sigma$ limiting magnitude & Seeing/PSF FWHM\\
\hline

\textit{u} &  25.6  &  1.22"\\

\textit{g}   & 26.8  &  1.09"   \\

\textit{r}  &  26.9 & 1.02"  \\

\textit{i}  &   26.4 & 0.99"  \\
\textit{z}  &  25.8 & 1.01"  \\

\textit{y}  &  24.8 & 0.99"  \\

Y106 &  26.6  &0.220"   \\

J129  & 26.6  &  0.231"   \\

H158  &  26.6 & 0.242"  \\

F184 &   26.0 & 0.253"  \\

W146 &  27.2 & 0.238"  \\

\hline
\end{tabular}
\end{adjustbox}
\caption {The expected 5$\sigma$ point source limiting magnitudes and PSF FWHM for LSST (\textit{ugrizy}) and \textit{Roman} (\textit{YJHFW}) coadded images. Values for LSST are for year 4. These parameters are used in \texttt{RomanErrorModel} and \texttt{LSSTErrorModel}, respectively, to produce semi-realistic observed magnitudes at the catalog level. The displayed PSF FWHM values for LSST are the default values used in \texttt{PhotErr} and are shown here for reference.}
\label{tab: phot_errmodel}
\end{table}

\begin{figure*}
    \includegraphics[width=0.35\linewidth]{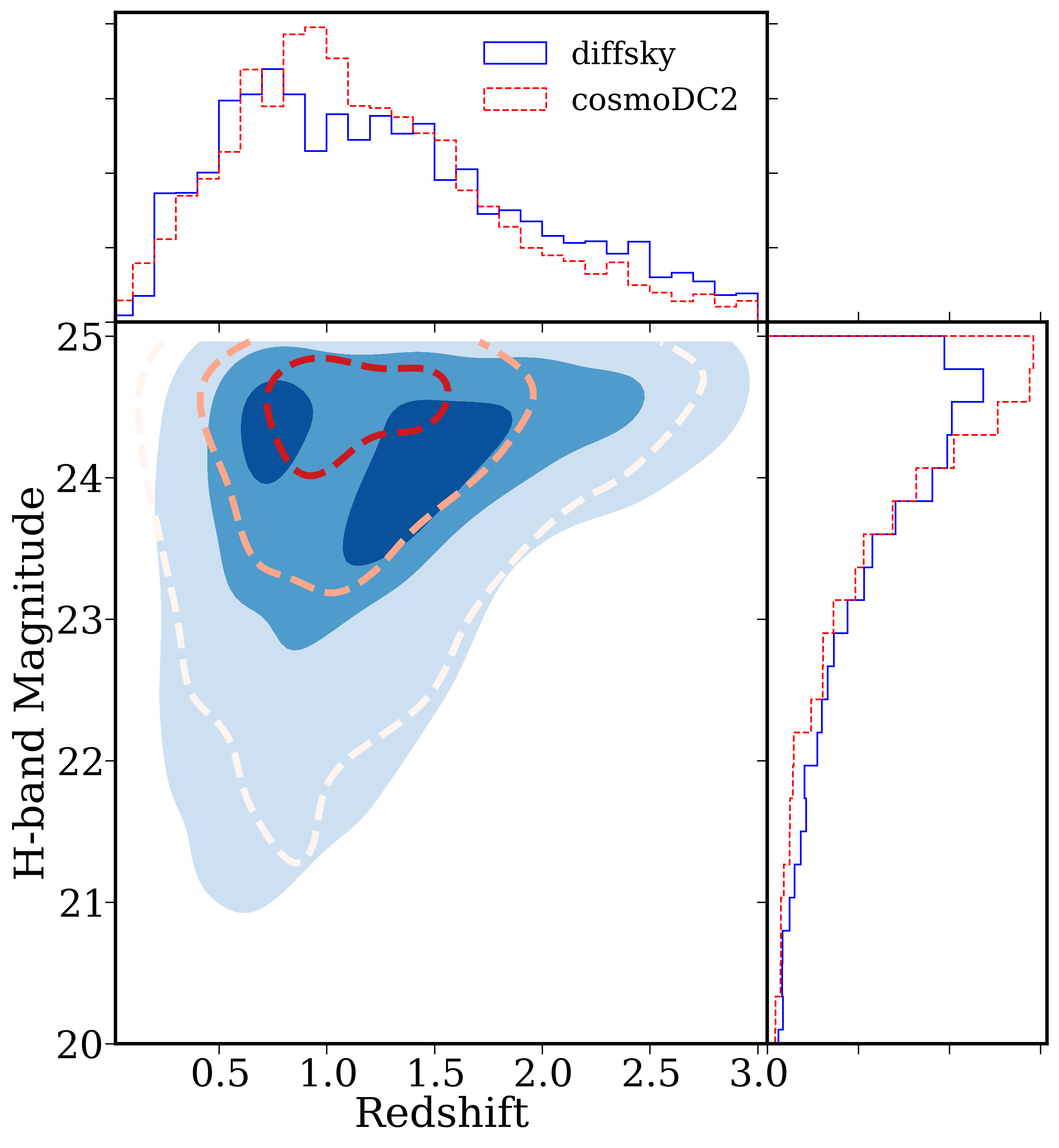}
    \hspace{0.5cm}
    \includegraphics[width=0.6\linewidth]{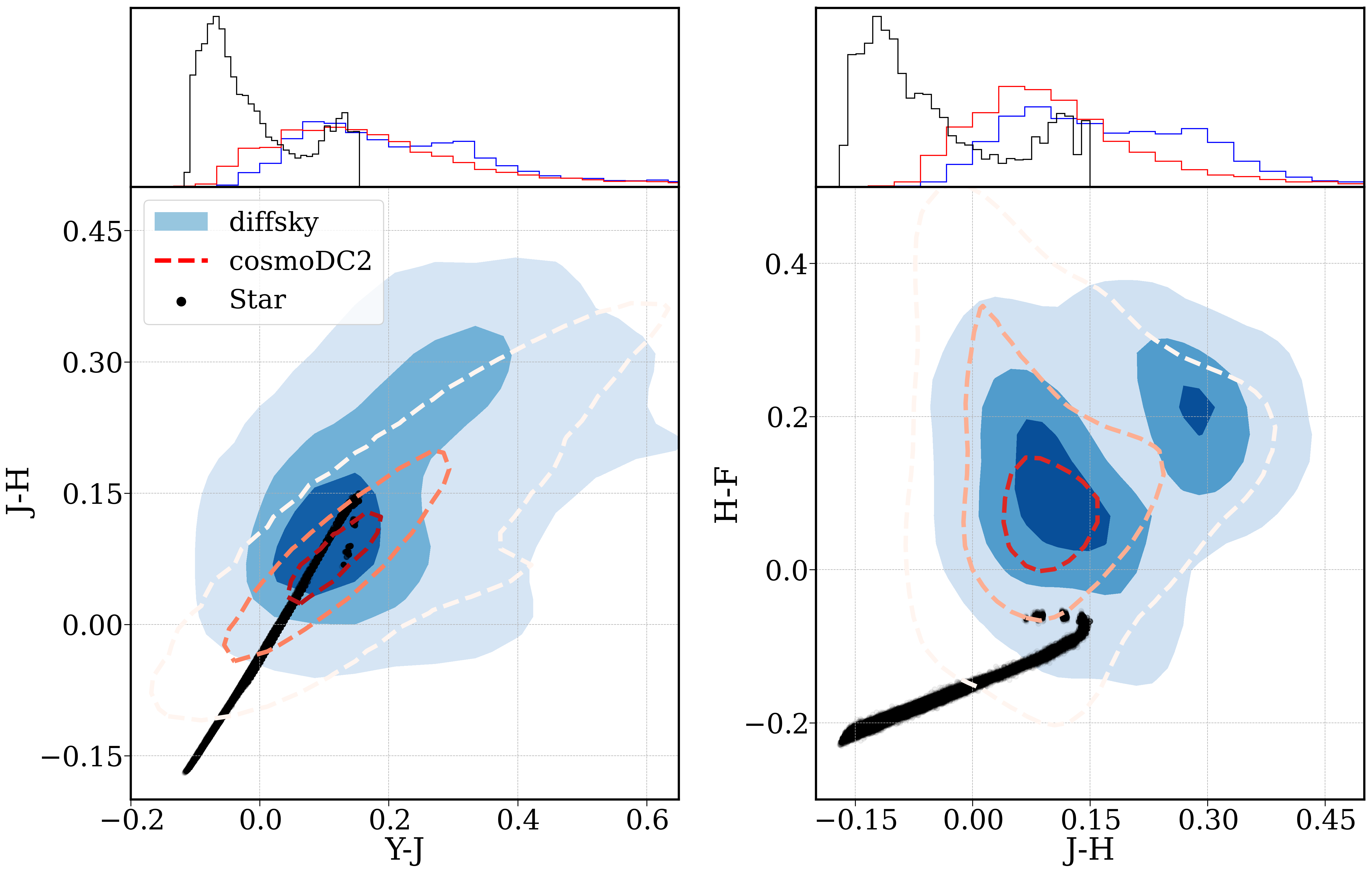}
    \caption{\textbf{Left}: Contour plot of the redshift vs.\ $H$-band magnitude for both \texttt{Diffsky} (filled blue) and cosmoDC2 (dashed red) galaxies. The adjacent 1D histograms show the respective distributions. A WL selection cut of $H < 24.96$ is applied to exclude galaxies with SNR $ <18$. All magnitudes shown are the observed quantities after applying the catalog-level noise described in Sec.~\ref{Simulations:noise}. We see that cosmoDC2 contains a higher number of faint objects in comparison to \texttt{Diffsky}. \textbf{Middle and Right}: The color-color plots for the galaxies (contours) and stars (black points) for the observed colors: $Y-J$, $J-H$, and $H-F$. The 1D histograms show the normalized color distributions for the x-axis. We observe multiple differences between the galaxies and stars in color space.  The stars not only occupy a narrower region of color space, but they also do not completely overlap the galaxy color distribution. These color differences will translate into differences in the effective PSF, and therefore bias shape measurement. All contours show 16th, 50th, and 84th percentiles of the data distribution.
    }

    \label{fig:sample_dist}
\end{figure*}

\subsection{Catalog-level Noise}\label{Simulations:noise}
We generate catalog-level noise to add to the true magnitudes in the galaxy and star catalogs in order to recreate semi-realistic conditions for the observed \textit{Roman} magnitudes. We also produce observed magnitudes for LSST bands, as these will be used later in this work. We use the photometric error model in \texttt{PhotoErr v1.3} \citep{crenshaw2024}, as it provides a simple way to generate noisy magnitudes and already has built-in models for both \textit{Roman} and LSST.

We will now detail the input parameters needed for the \textit{Roman} and LSST models, with text in parentheses showing the parameter names used in the code. The \texttt{RomanErrorModel} requires us to specify the 5$\sigma$ limiting magnitude (\texttt{m5}) and median zenith seeing FWHM\footnote{While the concept of zenith seeing is not relevant for \textit{Roman}, we provide the expected PSF FWHM so as to be able to use \texttt{PhotoErr}.} (\texttt{theta}) for each band.  In addition, we set the airmass (\texttt{airmass}) to 1 (setting to 1 denotes no airmass in this context) and specify the source type (\texttt{extendedType}) as `auto' for galaxies or `point' for stars. All other parameters are set to their built-in default values. The \texttt{LSSTErrorModel} is more developed than the \textit{Roman} one, and only requires us to provide the 5$\sigma$ limiting magnitudes, source type, and number of visits per year (\texttt{nVisYr}). We set
 \texttt{ nVisYr} = 1, as this allows us to provide the limiting magnitude for any year in LSST and bypass the internal calculation done by the code. Given the expected launch dates and mission lifetimes for \textit{Roman} and LSST, we chose to use the LSST limiting magnitudes for year 4 (Y4).

Table~\ref{tab: phot_errmodel} shows the input limiting magnitudes and PSF FWHM values used for each error model. All limiting magnitudes are for point-like sources. The values used in the case of \texttt{extendedType = `auto'} are 1 magnitude shallower for \textit{Roman} and 0.6 magnitudes shallower for LSST. This is an approximate relationship between the limiting magnitudes for extended objects and point-like sources for space- and ground-based telescopes like \textit{Roman} and LSST. The \textit{Roman} limiting magnitudes come from a DRM-like scenario for \textit{Roman}'s Wide Field Instrument (WFI). The LSST limiting magnitudes come from \texttt{v3.6} of the baseline survey strategy\footnote{Values available at \url{https://github.com/lsst-pst/survey_strategy/blob/main/fbs_3.6/v3.6_Update.ipynb}}. The provided values are for LSST Y10, so the values for Y4 depth are calculated using equation:
\begin{equation}\label{eq: lsst_depth}
\text{depth(year)} = \text{depth(Y10)} + \frac{2.5}{2} \log_{10} (\text{year}/10).
\end{equation}
This equation assumes that depth is limited only by the integration time. 
The median PSF FWHM is taken from \citet{Laliotis_2024}, which are used in the \textit{Roman} coadds \citep{Cao_2024} from the OpenUniverse image simulations \citep{OU_2025}, and is used here as the expected PSF FWHM for oversampled images.
PSF FWHM values are only reported for $YJHFK$ bands, and roughly follow a linear relationship with effective wavelength. We therefore estimate the value for the $W$ filter by interpolating using its effective wavelength. Further details on our useage of \texttt{PhotErr} are discussed in Appendix~\ref{appendix:mag_err}.


We acknowledge that our use of \texttt{PhotErr} violates multiple assumptions about \textit{Roman}, mainly the assumptions of a Gaussian PSF, which is not true for the native \textit{Roman} PSF.  It is also over-simplified in its use of an elliptical galaxy light profile and its neglect of blending and instrumental effects. We nevertheless believe that the level of realism provided is sufficient for the purposes of this work.




\subsection{Selection Cuts}
\label{subsec:selection-cuts}
 Two selection cuts are applied at the catalog level to remove objects that would not be used for real shear measurement analysis. The signal-to-noise ratio (SNR) of galaxies and stars in the catalog is calculated directly from the catalog-level noise described earlier. We apply a nominal SNR $ > 18$ cut for galaxies suitable for WL (see the \textit{Roman} Science Requirements Document\footnote{\url{https://roman.gsfc.nasa.gov/science/docs/RST-SYS-REQ-0020D_DOORs_Export.pdf}}, or SRD, for details), which roughly corresponds to a cut at $H < 24.96$. After this cut, we remove the few galaxies (fewer than 0.1\%) for which \texttt{PhotErr} returns observed magnitudes, in any filter, with a value of infinity. For stars we include only those bright enough for PSF modeling, so we exclude those with $\text{SNR} < 100$. Additionally, saturation cuts are applied to avoid issues with overly bright stars; specifically, we ensure that the brightest pixel in any star image is less than half the \textit{Roman} detector saturation level. These cuts will vary by filter, meaning the set of stars drawn will also vary between filters.

 Fig.~\ref{fig:sample_dist} shows the redshift-magnitude distribution (left plot) and the color-color distributions (right plot) for visualization of the final sample. The redshift-magnitude plot shows a larger number of faint objects for cosmoDC2 in comparison to \texttt{Diffsky}, but besides that the two distributions are fairly similar. The color-color plots display the differences between stars and galaxies in color space, with stars occupying a much smaller and not always overlapping region with galaxies. This difference motivates the need to study the impact of chromatic effects on the PSF and shear measurement.


\begin{figure*}
    \includegraphics[width=0.9\linewidth]{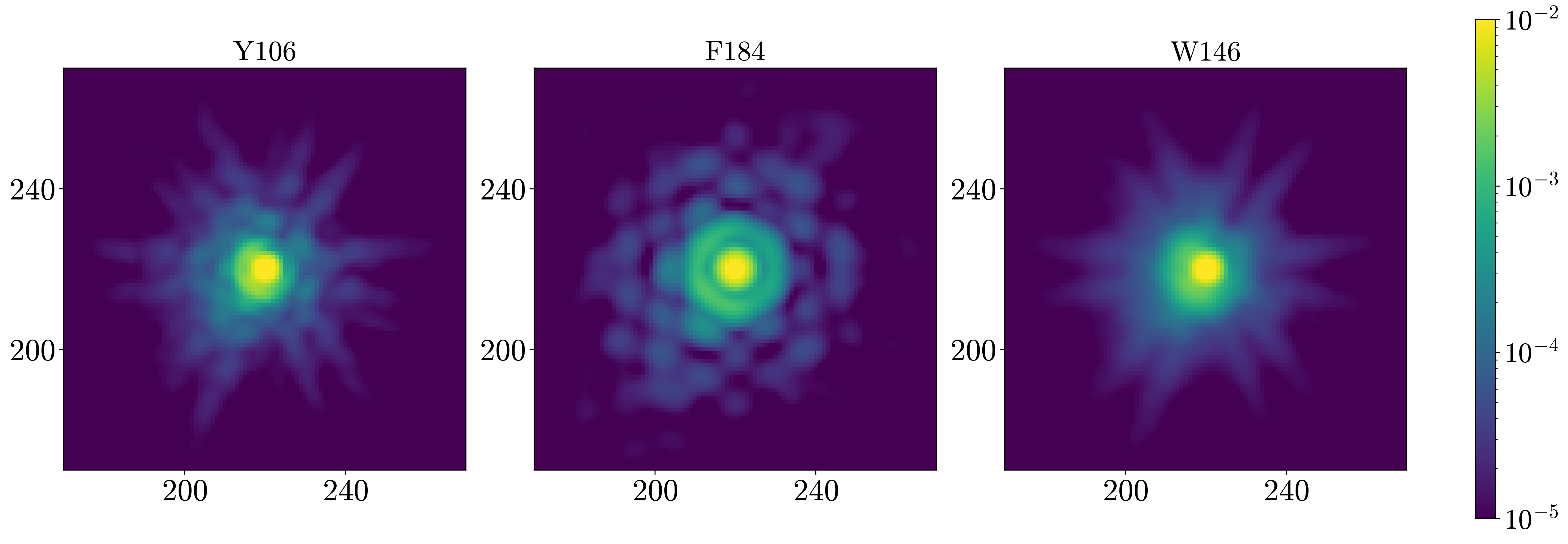}
    \caption{Examples of the normalized and oversampled \textit{Roman} PSF produced using the \texttt{galsim.roman} module for a random star in the catalog. From left to right we show the bluest (Y106) and reddest (F184) filters used for this analysis, and the wide filter (W146) PSF, simulated at a pixel scale of $ 0.0275 \text{ arcsec/pixel}$, one-fourth of the native \textit{Roman} pixel scale. We can visually see an increase in PSF size as we go from bluer to redder filters, which is expected for a diffraction-limited telescope like \textit{Roman}.
    }

    \label{fig:example_psf}
\end{figure*}

\subsection{Image Simulations using \texttt{GalSim}}\label{subsec:sims}

Image simulations are done using \texttt{GalSim v2.6} \citep{Galsim}, 
a widely-used Python package for simulating astronomical images. We specifically use the \textit{Roman} module within \texttt{GalSim}, which includes detailed observatory and instrument parameters, including PSF models.  
For these simulations, the \textit{Roman} PSF for any filter and position on the focal plane can be obtained by calling the \texttt{getPSF} function. No detector effects are 
included in these simulations. We simulate noiseless images of galaxies, stars, and the true galaxy effective PSF, in the standard \textit{Roman} WL filters ($YJHF$) and the wide filter $W$. The $Z$ and $K$ filters are excluded from our analysis due to the expected undersampling in the $Z$-band and low SNR in the $K$-band for the galaxies used for weak lensing. We use noiseless images to isolate biases due only to chromatic effects. In addition, the use of noisy images would require more simulated galaxies in order to beat down the statistical error. 

\begin{figure*}
    \includegraphics[width=0.95\linewidth]{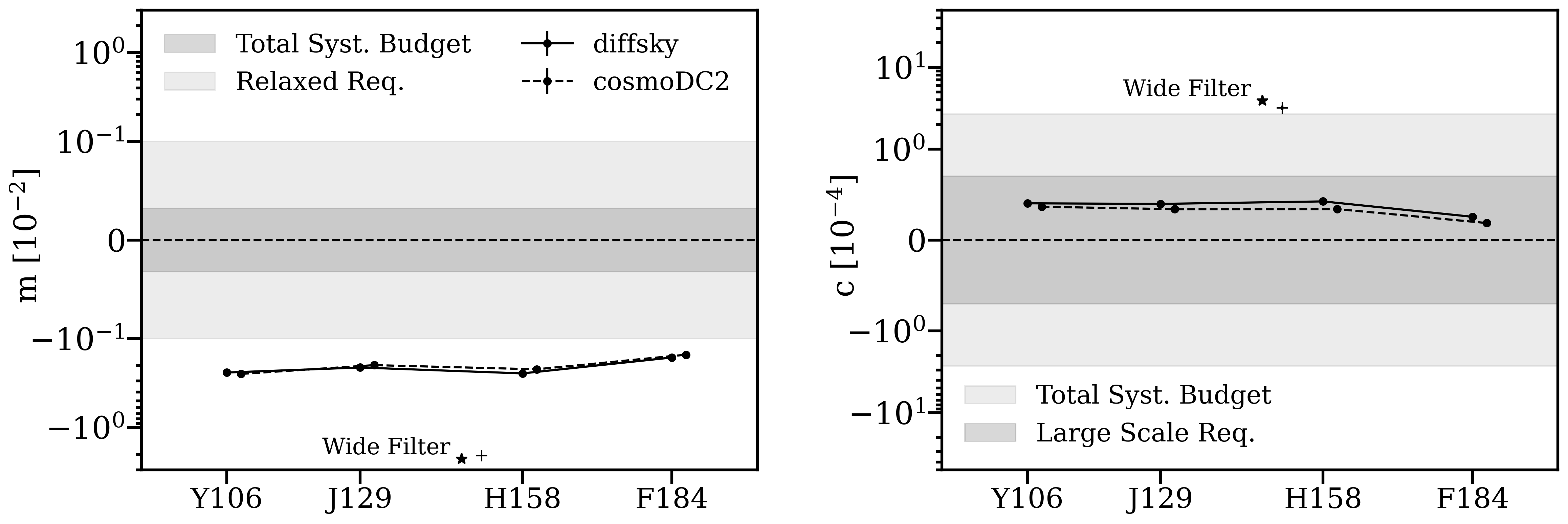}
    \caption{Shear measurement bias due to chromatic PSF effects averaged over all 10,000 simulated galaxies when no attempt is made to mitigate the effect.
    Error bars, calculated using a bootstrap method, are very small and may be challenging to see: for context, the typical uncertainty on $m$ is $4\times 10^{-5}$ for the WL bands and $3\times 10^{-4}$ for the wide filter. Uncertainties on $c$ are approximately an order of magnitude smaller than those on $m$. The statistical uncertainties are subdominant to the systematic uncertainties, which are not captured here. Results using the \texttt{Diffsky} 
    catalog are shown as solid lines, while those from cosmoDC2 are shown with dashed lines and a small horizontal shift for ease of visualization. \textbf{Left:} The multiplicative bias, expressed as a percentage, is shown for all 4 WL bands (points) and the wide filter (star for \texttt{Diffsky}, plus sign for cosmoDC2). The dark grey filled area represents the total multiplicative systematic budget for \textit{Roman}, while the light grey area represents a relaxed requirement of $|m| < 0.001$. \textbf{Right:} The additive bias is similarly shown for all 4 WL bands and the wide filter. The light grey filled area represents the total additive systematic budget for \textit{Roman}, obtained via the residual sum of squares of all scale bins to be used. The dark grey area indicates the most stringent requirement, which applies at large scales. As we can see from the left panel, 
    the multiplicative bias hovers around 0.2\% for the WL bands and 2\% for the wide filter when averaged over all galaxies. These biases exceed both a relaxed and stringent requirement. The additive biases (right panel) for the WL bands lie within the most stringent large scale requirement and hover just outside of the total systematic budget for the wide filter. Therefore, chromatic effects appear more problematic for multiplicative shear biases than additive biases. Biases for WL bands are roughly one order of magnitude smaller than those for the wide filter, highlighting possible problems with measuring shear using the wide filter. }
    \label{fig:uncorr_bias}
\end{figure*}

From the selection-cut catalog described in Sec.~\ref{subsec:selection-cuts}, 10,000 galaxies are selected at random and are placed at random positions in the focal plane. Each galaxy is paired with 40 unique stars (chosen from the subset prior to the star selection cuts), also selected at random. These represent a common set of stars for galaxies across all filters.  
The stars associated with each galaxy use the PSF at the same focal plane position as their respective galaxy, in order to explore biases due to chromatic effects without the need for spatial interpolation. The stellar PSF associated with each galaxy is determined as the flux-weighted average of the stars that meet the selection criteria outlined in Sec.~\ref{subsec:selection-cuts}. A total of 10,000 galaxies and 400,000 stars are used from the catalog to create these simulations. We confirm that our quantitative estimates of PSF errors due to differences in galaxy and stellar SEDs converge within 1\% after $\sim$1500 galaxies, ensuring that a sample of 10,000 is not subject to large sampling errors. To eliminate shape noise, we draw four versions of each galaxy with 45$^\circ$ rotations. For fitting the multiplicative and additive biases, galaxies are drawn with three different shears: $g_1 = \{-0.02, 0, 0.02\}$ and $g_2 = 0$.

We want to simulate Nyquist-sampled images for accurate shear measurement. Therefore, objects are not drawn at the undersampled native \textit{Roman} pixel scale (0.11 arcsec/pixel), but rather at one-fourth of the native pixel scale (0.0275 arcsec/pixel). This particular value is guided by the output scale in the coadded images in \cite{Hirata_2024} but is somewhat arbitrary. In order to create oversampled images and get the correct effective PSF, we first convolve all objects with the native pixel response function and then draw at the oversampled scale using the Galsim drawing method \texttt{no\char`_pixel}. This ensures that the object drawn is not convolved with the oversampled pixel response function, but is still drawn at the desired scale. The stamp size for all images is 440$\times$440 pixels, or roughly 12$\times$12 arcsec, to ensure we capture the entire light profile of larger galaxies. Fig.~\ref{fig:example_psf} shows an example of the normalized simulated effective PSF for the bluest (Y106) and  reddest (F184) filters used for our analysis, and the wide filter (W146). The image is zoomed in to capture the center of the PSF. We see clear differences between all images, highlighting the variability of the PSF with wavelength. We measure the PSF FWHM for each filter to be: Y106: 0.131\arcsec, J129: 0.139\arcsec, H158: 0.153\arcsec, F184: 0.169\arcsec, W146: 0.148\arcsec. Note that these values vary significantly from those in Table~\ref{tab: phot_errmodel}. The values reported in Table~\ref{tab: phot_errmodel} come from the coadded PSF in OpenUniverse2024, while the ones measured in our simulation come from direct oversampled images of the individual exposure PSF. Coaddition for \textit{Roman} in OpenUniverse2024 was done with PyIMCOM \citep{Cao_2024}, which uses a target PSF size larger than that of the true oversampled PSF of in individual exposures. The values in Table~\ref{tab: phot_errmodel} are only used to calculate magnitude errors.  



\subsection{Analytical Shear Estimation with \texttt{AnaCal}}

In this subsection we provide an overview of the Fourier Power Function Shapelets (\texttt{FPFS}) shear estimator introduced in \cite{Li_2018, Li_2022b} and implemented in the \texttt{AnaCal} framework \citep{Li_2023, Li_2024b}, which is used throughout this work for shape measurement. \texttt{AnaCal} is equipped to address various real-world observational challenges, including galaxy detection, selection, and noisy data. However, since the images used in this work are noiseless and galaxy detection is forced to avoid detection biases and focus solely on chromatic biases, we utilize only \texttt{AnaCal}'s shape measurement functionality for isolated, noiseless galaxies. Consequently, we will review only the aspects of AnaCal relevant to this study, with extensions to these simplified assumptions discussed in detail in \cite{Li_2023} and \cite{Li_2024a}.

Galaxy shear can be modeled as a locally linear transformation described by the Jacobian matrix:
\begin{equation}
A =
\begin{pmatrix}
1 - \gamma_1 & -\gamma_2 \\
-\gamma_2 & 1 + \gamma_1
\end{pmatrix},
\end{equation}
where $\gamma_1$ represents horizontal or vertical stretching, and $\gamma_2$ corresponds to stretching at 45-degree angles with respect to the horizontal direction. As described in Sec.~\ref{background:Wl and shear bias}, the ensemble average of galaxy ellipticities is used to estimate the shear. The ensemble weak-lensing shear, $\gamma_\alpha$, is estimated as:
\begin{equation}\label{eq:ensemble_shear}
\hat{\gamma}_\alpha = \frac{\langle e_\alpha \rangle}{\langle R_{\alpha} \rangle},
\end{equation}
where $e_\alpha$ is the galaxy ellipticity for component $\alpha$, and $R_{\alpha}$ (referred to as the shear response) is the partial derivative of the ellipticity with respect to $\gamma_{\alpha}$. Chromatic effects primarily impact the observed ellipticities $e_\alpha$, because errors in the estimated PSF size can bias the observed shape of galaxies. Both quantities on the right hand side equation~\eqref{eq:ensemble_shear} are estimated in \texttt{AnaCal} by calculating the weighted polar shapelet modes, $M_{nm}$ \citep[see equation~28 in][]{Li_2023}, which depend on the galaxy image, PSF image, and the polar shapelet basis \citep[see equation~25 in][]{Li_2023}. Note that in order to calculate these modes, \texttt{AnaCal} requires the user to specify a size, $\sigma_h$, for the polar shapelet basis. Typically, $\sigma_h$ is selected to be greater than the PSF's scale radius in real space \citep{Li_2023}. In this work we set $\sigma_h$ to be 15\% larger than the PSF scale radius of the particular filter in which the measurement is being made.

The galaxy ellipticities, $e_1$ and $e_2$, are then calculated:
\begin{equation}
e_1 = \frac{M_{22c}}{M_{00} + C}, \quad e_2 = \frac{M_{22s}}{M_{00} + C},
\end{equation}
where $C$ is a weighting parameter introduced in \cite{Li_2018} to stabilize the measurement by balancing contributions from galaxies of varying brightness levels and suppressing higher-order noise terms. In this work we set $C = 10$, which is close to the optimal value for LSST \citep{Li_2024b}. Since our images are noiseless, we only care about setting a reasonable value for $C$ that avoids giving very large weights to the brightest galaxies. We do not attempt to find the optimal value here, but in real applications it should be optimized for the \textit{Roman}-specific zero point and noise level. 
The shear response is also calculated using the shapelet modes described here \citep[see Eqs.~38 \& 39 in][]{Li_2023}.

We emphasize that the estimator described here relies on several simplifying assumptions: no shear-dependent detection/selection, shear-dependent weighting, or image
noise. These assumptions are appropriate for this work, but corrections for detection and selection biases, as well as noise, are necessary for shear estimation with real data and can be carried out using \texttt{AnaCal}.

\section{Size of Chromatic Bias}\label{Bias}

In this section, we present the shear measurement biases caused by chromatic PSF effects due to the differences between galaxy and stellar SEDs for 10,000 simulated galaxies across the four \textit{Roman} WL bands and the wide filter. All biases discussed correspond to the shear component \(\gamma_1\), as a shear is applied only in this direction in our simulations. Similar biases are expected in \(\gamma_2\). We quantify the multiplicative and additive biases, \(m\) and \(c\), using equation~\eqref{eq:obs_shear}. Shear is estimated using equation~\eqref{eq:ensemble_shear}, where the galaxy ellipticities and shear responses are calculated using the stellar PSF associated with each galaxy (see Sec.~\ref{subsec:sims} for details).

Figure~\ref{fig:uncorr_bias} illustrates the multiplicative (left plot) and additive (right plot) biases for the standard WL and wide filters for the galaxy ensemble in our simulations. We show this for both extragalactic catalogs, \texttt{Diffsky} and cosmoDC2. The left plot includes two bands representing systematic requirements: the narrower band corresponds to \textit{Roman's} multiplicative shear bias mission requirement, \(|m| < 3\times 10^{-4}\), as per the Roman~SRD, 
and the wider band corresponds to a requirement of \(|m| < 0.001\), closer to requirements for LSST and \textit{Euclid} \citep{LSST_2018, Euclid_2019}. 
The right plot includes two bands representing systematic requirements from the \textit{Roman}~SRD: the narrower band corresponds to \textit{Roman's} most stringent requirement (at large scales), while the wider band corresponds to the total additive systematic budget for \textit{Roman}, derived from the residual sum of squares across all scale bins. We note that the amplitude of systematic biases in shear estimation is sufficiently small when using the galaxy effective PSF—constructed from the galaxy's composite SED—such that the biases shown in Figure~\ref{fig:uncorr_bias} can be attributed entirely to chromatic effects, rather than to third-order shear biases or other issues in the shear estimation process. We discuss the implications of using the galaxy composite SED to construct the galaxy PSF and its implications on galaxy color gradients in Sec \ref{color_gradients}.

The results, which are almost identical between \texttt{Diffsky} and cosmoDC2, reveal a  multiplicative shear bias of approximately 0.2\% for the WL filters and about 2\% for the wide filter when averaged over all galaxies. We note that the bias can vary with redshift, with some redshift bins (not shown) reaching biases between 0.4-- 0.9\% for the WL bands and 3-- 6\% for the wide filter.  Thus, chromatic biases exceed both \textit{Roman} and general WL multiplicative bias requirements in all filters, and are especially egregious for the wide filter. Conversely, additive biases in the WL filters lie within the large-scale requirements, making them less problematic. However, additive biases in the wide filter exceed the total systematic budget, posing additional challenges for this band. Regardless of filter choice, these results show that mitigation methods are needed to correct for multiplicative biases if we wish to stay within survey requirements, while additive biases are only an issue for the wide filter.

It is important to note that \textit{Roman's} requirements budget as shown on the plot is for all systematics affecting shear estimation, not just chromatic effects. If we wish to restrict chromatic biases to meet a relaxed multiplicative bias requirement of \(|m| < 0.001\), the correction must be accurate (on average over all galaxies) to the 50\% level for WL bands and 95\% for the wide filter. If the aim is to stay within the dark grey region, the correction must be accurate to the 85\% and 99\% level for the WL and wide filters, respectively. Achieving this level of precision for the wide filter, in either case, is particularly challenging due to our limited knowledge of real galaxy SEDs and the complexity of modeling chromatic aberrations in the PSF across a broad wavelength range. This reinforces the great difficulty of correcting for chromatic effects in this band, and therefore the intrinsic challenge of accurate shape measurement in a potential wide filter for \textit{Roman}.

\section{Mitigation Method}\label{Mitigation}

\subsection{Chromatic Difference Formalism}\label{Mitigation:chrom_diff}

Given that the size of the uncorrected chromatic bias exceeds both the \textit{Roman} requirements and even a relaxed requirement of $|m| < 0.001$, methods to mitigate the chromatic biases are needed. We choose to work on PSF-level correction methods rather than correcting for biases after shape measurement. The motivation behind this choice is two-fold. The first is that PSF-level corrections are independent of the shape measurement method, meaning that they can be applied for any shear estimation algorithm of choice. The second is that for \textit{Roman} we expect to have a very accurate model of the PSF, and we hope to use that model to build an accurate chromatic PSF mitigation method.

We start from equation~\eqref{eq:eff_psf_diff}. Assuming that the position- and wavelength-dependence of the PSF is separable, 
we can Taylor expand the flux-normalized SED of some object $o$ around the filter effective wavelength, $\lambda_0$:
\begin{align}\label{eq:sed_approx}
     &S_{o}(\lambda) = S_o^{(0)} + S_o^{(1)}(\lambda - \lambda_0) + ... + \frac{S_o^{(n)}}{n!}(\lambda - \lambda_0)^n,\\
     &\text{where} \; \; \lambda_0 = \frac{\int \mathrm{d}\lambda \; F(\lambda) \lambda}{\int \mathrm{d}\lambda \; F(\lambda)},\nonumber
\end{align}
and $S_o^{(n)}$ is the $n$th derivative of the SED evaluated at $\lambda_0$. The differences between the normalized SEDs of stars and galaxies can then be written as: 
\begin{align}\label{eq:sed_diff}
   &S_{\star}(\lambda) - S_{\text{g}}(\lambda) =  \sum_{n = 0}^{\infty} \Delta S_n (\lambda - \lambda_0)^n; \; \; \Delta S_n = \frac{S_{\star}^{(n)} - S_{\text{g}}^{(n)}}{n!}.
\end{align}
Plugging this into equation equation~\eqref{eq:eff_psf_diff} we get:
\begin{align}\label{eq:psf_diff_approx}
    \Delta \text{PSF}_{\text{eff}} (x,y) &\approx \sum_n \Delta S_n \int \mathrm{d}\lambda \;  \text{PSF} (x,y,\lambda)\; F(\lambda) \; (\lambda - \lambda_0)^n\\
    &\equiv \sum_n \Delta S_n B_n (x,y)\nonumber
\end{align}
where we can identify the integral as an SED-independent basis of images, which we define as $B_n$, and $\Delta S_n$ as the coefficients describing the chromatic error in the PSF model as a linear combination of the basis functions. To the leading order, the shear biases are linearly dependent on the error in the PSF model and therefore on the coefficients. 
This formulation is advantageous because the SED-independent basis only depends on the filter throughput and the PSF model, which means that it can be precomputed in simulations given an accurate PSF model. The only remaining task to correctly quantify the bias is to correctly estimate the coefficients $\Delta S_n$, which only depend on the object's SED and can be inferred/learned without
image simulations. However, this method is efficient only if the chromatic shear bias depends on a small number of terms, as constraining higher-order terms in practice could be extremely challenging.

Assuming a linear relationship between the SED and wavelength, $S_o \approx m_o(\lambda -  \lambda_0) + b_o$, would simplify equation~\eqref{eq:psf_diff_approx} to:
\begin{align}\label{eq:psf_diff_linear}
    \Delta \text{PSF}_{\text{eff}} (x,y) &= \Delta S_0 B_0(x,y) +  \Delta S_1 B_1(x,y)
\end{align}
However, if the star and galaxy SEDs are normalized by their respective fluxes, the first term in that equation goes to zero (see Appendix~\ref{appendix:linear_proof}) and we have a single term to estimate the chromatic bias:
\begin{align}\label{eq:psf_diff_linear_norm}
    &\Delta \text{PSF}_{\text{eff}} (x,y) \approx \Delta S_1 B_1(x,y).
\end{align}
What this means is that if the integral of the galaxy and star SEDs can be well-approximated by the integral of a linear function within the filter, the chromatic bias is captured entirely by a single basis function and the difference between the galaxy and star SED's flux-normalized slopes.  Since this statement is true at the integral level (as per Eq.~\ref{eq:psf_diff_approx}), it is true not only for strictly linear SEDs, but also for ones that could be approximated as linear plus fluctuations of opposing signs that cancel out in the integral.  Moreover, this need only be true at the ensemble level rather than for each individual galaxy.
%
Note that the SED linearity assumption is more likely to hold for narrow filters than for wider one and when the SED continuum—the smoothly varying part of the SED—lies within the filter's wavelength range. In contrast, when prominent features such as the Balmer break enter the bandpass, higher-order terms in the Taylor expansion may be required. Finally, the first-order corrected effective PSF is given by:
\begin{align}\label{eq:corrected_psf}
    \text{PSF}_{\text{eff}}^{\text{corr}}& = \text{PSF}_{\text{eff}, \star} (x,y) - \Delta \text{PSF}_{\text{eff}} (x,y) \\
    &\approx \text{PSF}_{\text{eff}, \star} (x,y) - \Delta S_1 B_1(x,y).\nonumber
\end{align}

Realistic galaxy SEDs are clearly not linear across a wide wavelength range, but the simplicity of this formalism motivates a test of whether this approximation is sufficient within the filters used for \textit{Roman}.


\subsection{Practical Implementation}

The formalism in Sec.~\ref{Mitigation:chrom_diff} separates the bias into a set of SED-dependent coefficients, $\Delta S_n$, and PSF-dependent basis functions, $B_n$ (which are images). In this section we will detail how both components are computed.

\subsubsection{PSF Basis Functions} \label{Mitigation:psf_basis}
The PSF-dependent basis functions, $B_n$, are given by equation~\eqref{eq:psf_diff_approx} and depend on the PSF model $\text{PSF} (x,y,\lambda)$, the filter throughput $F(\lambda)$, and the basis functions $(\lambda - \lambda_0)^n$. 
If we look at equation~\eqref{eq:psf_diff_approx}, the integral describing the basis can be thought of as the effective PSF of an object with a fixed nonphysical SED $(\lambda - \lambda_0)^n$. For a known filter throughput, the only challenge in computing $B_n$ is in modeling the chromatic, spatially dependent PSF. However, given that \textit{Roman} is a space-based telescope, we expect to be able to model the PSF extremely well from ray-tracing simulations, detector tests and characterization, and empirical calibration from real data. Therefore, we expect to have a reliable model for $\text{PSF} (x,y,\lambda)$ for \textit{Roman}. Substantial work to characterize the \textit{Roman} PSF has already been done and implemented in \texttt{GalSim} and \texttt{WebbPSF}, where a full model of $\text{PSF} (x,y,\lambda)$ is available. In this work we use the model provided by \texttt{GalSim}, as this is also the software used for image simulations.

Given $\text{PSF} (x,y,\lambda)$, we can then construct $B_n$ fairly easily using image simulations. In theory this could be done at every position in the focal plane, however this would be computationally expensive. Instead, we chose to compute $B_n$ only at the center of each Sensor Chip Assembly (SCA), leaving an approach to account for both spatial and chromatic dependence of the PSF for future work. So for a given a focal plane position, $(x,y)$, that falls within SCA $\alpha$, the PSF basis function is approximated as:
\begin{equation}
    B_n (x,y) \approx  B_n(x^{\alpha}_{\text{center}}, y^{\alpha}_{\text{center}}) ,
\end{equation}
where $(x^{\alpha}_{\text{center}}, y^{\alpha}_{\text{center}})$ is the focal plane position of the center of SCA $\alpha$. This decision to approximate $B_n$ assumes that spatial differences between the PSF model at different positions within a single SCA are negligible when correcting for chromatic biases. This makes sense, chromatic biases are a sub-percent (percent) level bias for the standard WL bands (wide filter), so a sub-percent level bias due to this spatial approximation would be smaller than the most stringent multiplicative bias requirements. This approximation is tested later on for $B_1$ and confirmed to hold.

Given the above, here is the recipe for simulating $B_n$ in any given filter using \texttt{GalSim}:
\begin{enumerate}
    \item Retrieve \( \text{PSF} (x,y,\lambda) \) using \texttt{galsim.roman.getPSF}.
    \item Obtain $F(\lambda)$ using \texttt{galsim.roman.getBandpasses}.
    \item Create a delta function object, \texttt{galsim.DeltaFunction}, and multiply it by an \texttt{SED} object defined as \( (\lambda - \lambda_0)^n \), where \( \lambda \) spans 1000 evenly spaced wavelengths between the filter’s blue and red limits.
    \item Set the SED units to \texttt{\detokenize{`fphotons'}} (photons/nm/cm\(^2\)/s) using the \texttt{flux\_type} argument. This is done to match \texttt{GalSim}'s output units of photons per pixel in the images.
    \item Convolve the \texttt{GalSim} object with the pixel response at the native \textit{Roman} pixel scale.
    \item Draw the final object at desired pixel scale (using \texttt{method = `no\_pixel'}).  Here we chose to draw at one-fourth of the native \textit{Roman} scale. The resulting image is $B_n (x,y)$ from equation~\eqref{eq:psf_diff_approx}.
\end{enumerate}


\begin{figure*}
    \includegraphics[width=0.95\linewidth]{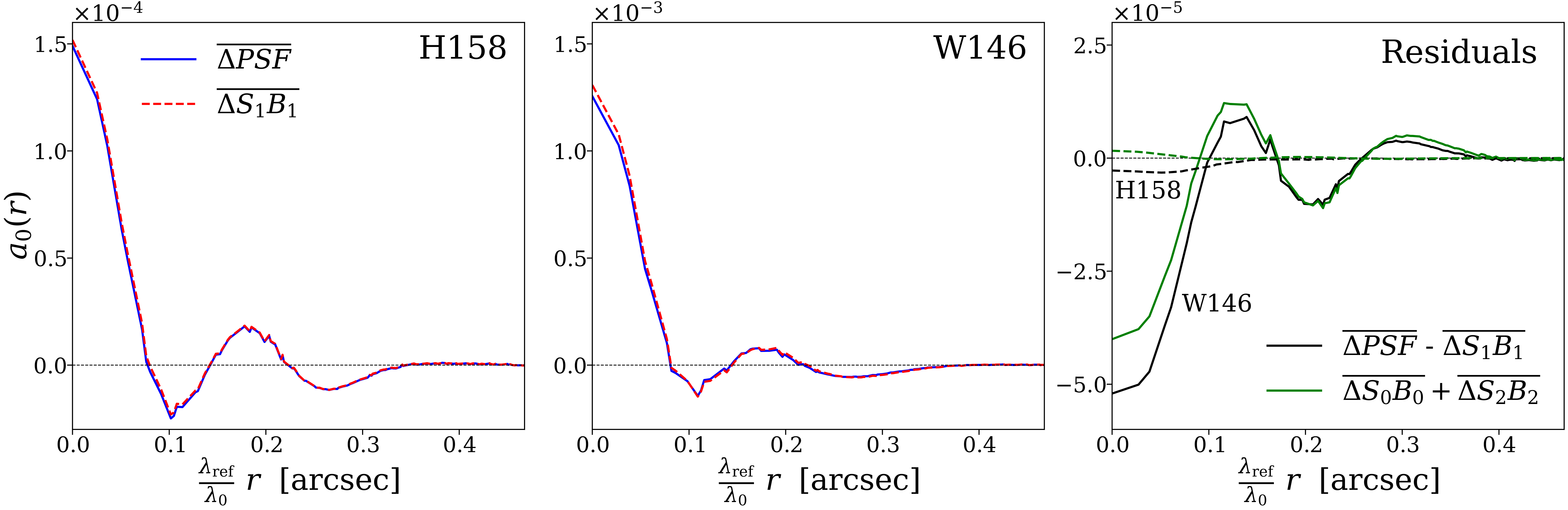}\\
    \caption{Comparison of the average spin-0 component radial profile, $a_0(r)$, of the chromatic PSF differences from image simulations using \texttt{Diffsky} (blue solid line) and the first-order term in the Taylor expansion (red dotted line). The x-axis is normalized by the ratio of a chosen reference wavelength, $\lambda_{\text{ref}} = 1460 $ nm, and the filter effective wavelength $\lambda_0$. $\overline{\Delta \text{PSF}}$ is calculated over all galaxy-star pairs, while the first-order approximation to the bias, $\overline{\Delta S_1 B_1}$, is derived from simulated PSF basis function and fitted SEDs as described in Secs.~\ref{Mitigation:psf_basis} and~\ref{Mitigation:SED_fit}. Results are shown for the H158 (left) and W146 (center) filters. Results for the other WL bands are similar to those for H158. Residuals between the observed bias and the first-order approximation for both filters are shown in the right panel as black curves (dashed for H158 and solid for W146). The near-overlap of the blue and red curves in the left and middle panels suggests that the first-order term predominantly explains the chromatic differences, on average. Upon closer inspection of the residuals in the right panel, this is true for H158 but not for W146. This can be understood by considering the difference in y-axis scale between the left and center panels. 
    The remaining bias in W146 can be explained by including the second-order terms (green) in the approximation (the $\Delta S_0$ term is non-zero if the SED is not linear). This suggests that the chromatic differences between galaxy and stellar PSFs can be effectively described using only the first-order term for the WL bands, but requires higher order terms for the wide filter W146.
    }

    \label{fig:basis}
\end{figure*}

\subsubsection{SED Coefficients}\label{Mitigation:SED_fit}

The SED coefficients, $\Delta S_n$, depend on the star and galaxy SEDs or the expansion coefficients for the flux-normalized SEDs, $S_{\star}^{(n)} \text{ and }\;  S_{\text{g}}^{(n)}$. With full knowledge of both the galaxy and star SEDs we can estimate these quantities by doing an $n^{\text{th}}$ degree polynomial fit to the SED within the desired filter. We have found empirically that given the coarse sampling of the SED in our catalogs, in order to properly estimate the coefficients that correct the chromatic biases in the images, the fit cannot simply be done by fitting the provided SED values within the red and blue filter limits. Instead, the SED must be interpolated to obtain a much finer spectral resolution, and the fit must be weighted by the filter throughput. That is, if we wish to estimate $S_g^{(1)}$ for the H-band given the true galaxy SED, we first interpolate the SED inside the filter to a higher spectral resolution (we use 1000 evenly space wavelengths within the filter), and then fit the fine-grained SED to a 1st degree polynomial, weighting each point in the fit by the filter transmission value corresponding to that wavelength. This is done for both the star and galaxy SEDs in order to obtain their difference, $\Delta S_1$. The SED interpolation is done using a natural cubic spline, \texttt{GalSim}'s default internal interpolation method. This form of interpolation uses a piecewise cubic polynomial to fit the data, ensuring smoothness by having the second derivative be continuous across the interval and zero at the endpoints.


\subsection{Mitigation Performance in the Perfect Scenario}


We evaluate the mitigation performance at the PSF and shear measurement level to demonstrate the effectiveness of the method when both $\Delta S_1$ and $B_1$ are known perfectly. Recall that the PSF used for shear inference is being directly corrected. With these assumptions, failures of the correction method are due to the inability of the linear SED model to capture the impact of the true SED shape within the filter.

We first compare the PSF residuals to the first-order approximation described previously. To do so, we compute the average PSF bias, $\overline{\Delta \text{PSF}}$, and average first-order approximation, $\overline{\Delta S_1 B_1 }$, for all simulated galaxy-star pairs. Since these are 2D images, we chose to quantify the effectiveness of the correction at the PSF-level by looking at the radial profile of the spin-0 component, $a_0 (r)$, defined as:
\begin{equation}
    a_0 (r) =\frac{1}{2\pi} \int f(r,\theta) \mathrm{d}\theta,
\end{equation}
where $f(r,\theta)$ is the 2D image of the quantity of interest, with $r$ and $\theta$ being the radial distance and azimuthal angle, respectively, from the center of the image. The radial profile of the $\overline{\Delta \text{PSF}}$ spin-0 component serves as a more detailed description of the average bias in the PSF size, and therefore the multiplicative bias on shear. The spin-2 component would resemble the equivalent but for PSF shape errors (connected to additive shear biases).  However, given the results in Fig.~\ref{fig:uncorr_bias}, these are a smaller concern than the multiplicative biases. Therefore, we only chose to explore $a_0$.

Fig.~\ref{fig:basis} compares $a_0 (r)$ for $\overline{\Delta \text{PSF}}$ (blue solid line) and $\overline{\Delta S_1 B_1 }$ (red dashed curve) for the H158 (left panel) and W146 (center panel) filters. Results for the other WL filters resemble those shown for H158. Results are shown only for the \texttt{Diffsky} catalog, but are similar for cosmoDC2. The right panel shows residuals (black curves) between the observed bias and first-order approximation for both filters (dashed for H158, solid for W146). We first note the order-of-magnitude difference in scale between the y-axis of the left and center panels. Since the amplitude of $\overline{\Delta \text{PSF}}$ is roughly proportional to the multiplicative bias, we can expect to see biases for the wide filter that are roughly an order of magnitude bigger than those of the other WL bands, consistent with Fig.~\ref{fig:uncorr_bias}.

An initial glance at the left and center panels would suggest that the first-order approximation suffices, on average, to estimate the PSF biases for all filters. However, residuals on the right plot show that this holds only for H158 (and the other WL bands not shown), but not for W146, as there is still a considerable amount of bias left. We can explain most of the remaining bias in W146 by looking at the additional terms (green curve) of a 2nd degree approximation. This suggests that more accurate correction of the chromatic biases for W146 would likely require a second-order approximation for the SED.

\begin{figure*}
    \includegraphics[width=0.9\linewidth]{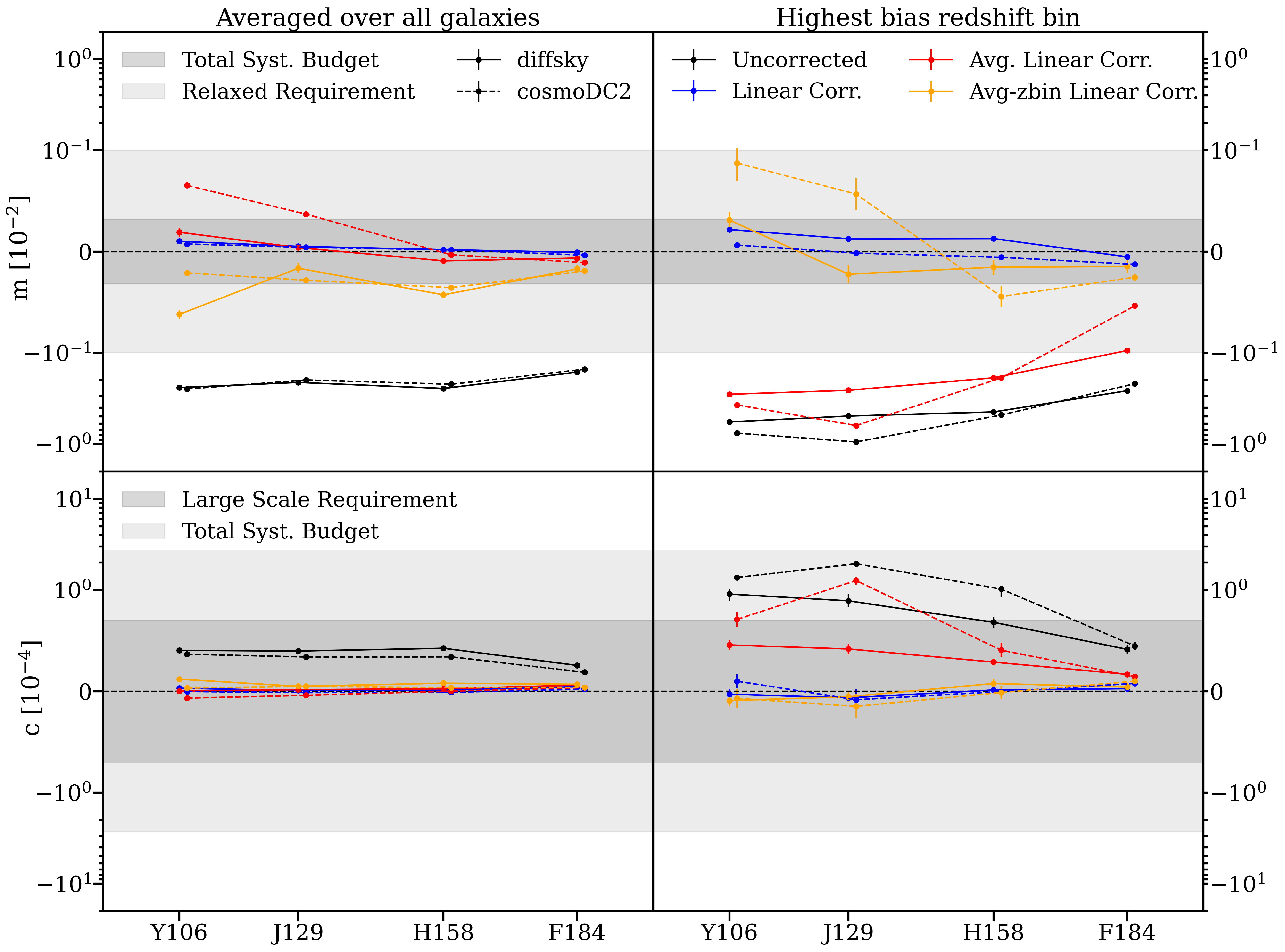}\\
    \vspace{0.4cm}
    \includegraphics[width=0.85\linewidth]{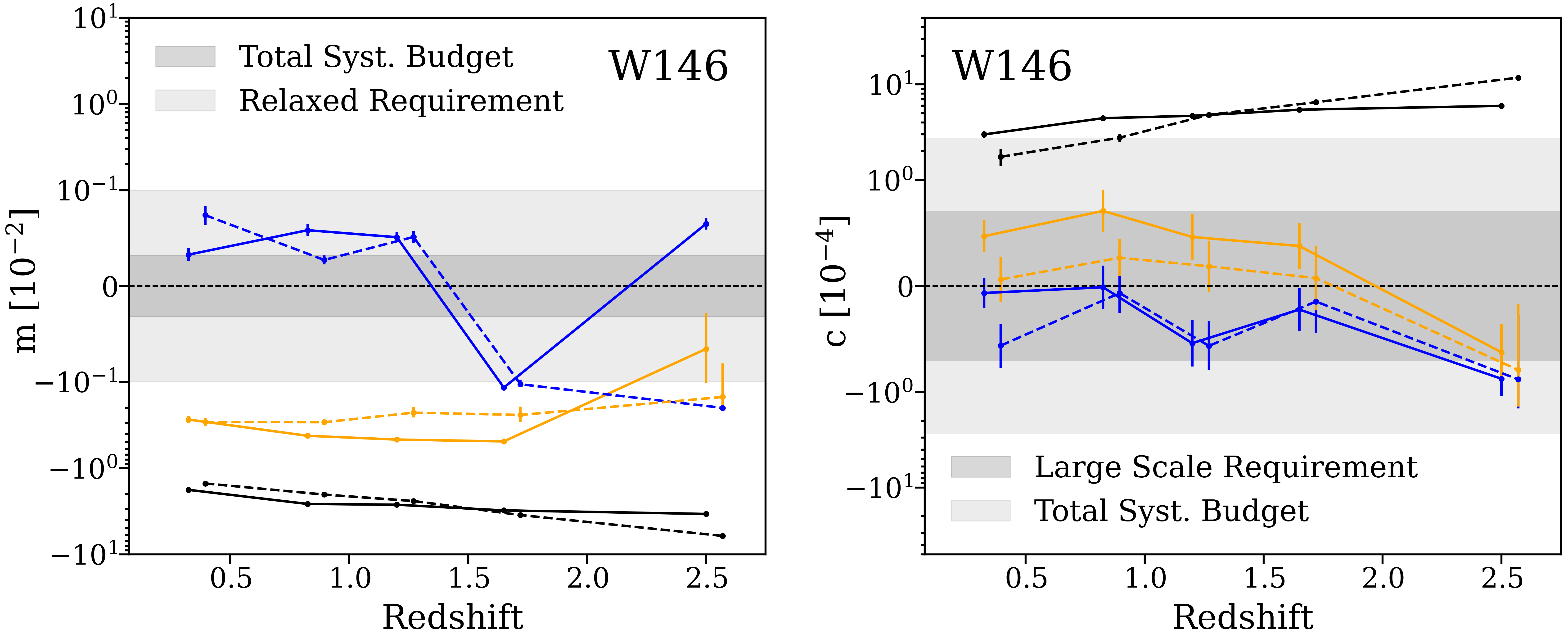}

    \caption{Shear measurement multiplicative and additive biases after applying image-level correction to the effective stellar PSF using the first-order linear correction for both \texttt{Diffsky} (solid lines) and cosmoDC2 (dashed lines). We test the per-galaxy correction using the true estimated values of $\Delta S_1$ (blue), the true average of $\Delta S_1$ over all galaxies (red), and the true average over all galaxies in each redshift bin (orange). The black curves show the uncorrected bias. The grey bands represent the requirements described in Fig.~\ref{fig:uncorr_bias}. \textbf{Top:} The multiplicative (additive) biases for the 4 WL bands in the top (bottom) panels. The left column shows bias calculated over all galaxies (similar to Fig.~\ref{fig:uncorr_bias}), while the right panel shows the bias for the worst-performing redshift bin for each filter. Using the true $\Delta S_1$ for each galaxy reduces biases to meet stringent multiplicative bias requirements, demonstrating that the first-order approximation is sufficient when parameters are accurately estimated. While corrections based on the average galaxy SED do not meet the relaxed requirements at certain redshift bins, the correction using redshift bin-averaged SED coefficients safely meets the relaxed requirements. \textbf{Bottom:} The multiplicative (left) and additive (right) biases as a function of redshift for the wide filter. The red curve is not shown due to its under-performance, completely failing to correct for biases. The perfect first-order approximation does not meet the strictest multiplicative bias requirements, and for some redshift bins, even the relaxed requirement. We note that even though the correction does not look too bad for W146, it fails to converge and can be unstable, underscoring the challenge of accurate shape measurements with the wide filter even with perfect knowledge of the PSF and SED. 
    }

    \label{fig:corr_bias}
\end{figure*}

Fig.~\ref{fig:corr_bias} now shows the performance of the linear chromatic mitigation method on shear measurement using \texttt{AnaCal}. We test the per-galaxy correction using the true estimated values of $\Delta S_1$ (blue curve), the true average of $\Delta S_1$ over all galaxies (red curve), and the true average for each redshift bin (orange curve). `True' here refers to the values measured from the galaxy and star SEDs. In all cases the PSF correction is done per-galaxy, the only difference between the cases described is what prefactor goes in front of the $B_1(x,y)$ term in equation~\eqref{eq:corrected_psf}. 
We show the results for both \texttt{Diffsky} (solid lines) and cosmoDC2 (dashed lines). The top (bottom) left column shows the multiplicative (additive) bias calculated over all galaxies (similar to Fig.~\ref{fig:uncorr_bias}), while the right panel shows the bias for the worst-performing redshift bin for each filter. The two bottom panels show the multiplicative and additive biases for the wide filter as a function of redshift.

Looking at the results for the WL bands, we can clearly see that using the true $\Delta S_1$ for each galaxy mitigates the biases to within the most stringent multiplicative bias requirements, showing that the first-order approximation is sufficient when these parameters can estimated accurately. The correction based on the average galaxy SED 
performs well when looking at the total ensemble (as expected), but can fail to stay within a relaxed requirement at certain redshift bins. On the other hand, the correction using the SED averaged within individual redshift bins 
can remain within a relaxed requirement, but not within the stricter requirement from the Roman SRD. These results show that if the goal is to mitigate chromatic biases to the strictest requirements, an accurate per-galaxy correction is needed. On the other hand, if staying within a relaxed requirement is enough, accurately estimating the average correction for each tomographic bin might suffice. We note that these conclusions do not vary much between \texttt{Diffsky} 
and cosmoDC2. The results for the additive biases show that all correction methods can safely stay within the dark gray region, with the exception of the overall average method for Y106 and J129 when using the cosmoDC2 catalog.

Looking at the bottom panels for the wide filter, we see a different story. The first-order approximation fails to stay within the strictest multiplicative bias requirements, and for some redshift bins, even the relaxed requirement. We note that even though the correction does not look too bad for W146, it fails to converge. Doing the same correction using a second-order polynomial fit to the SED (which is meant to be more accurate) resulted in a higher multiplicative bias, failing to stay within the relaxed requirement in all redshift bins for both extragalactic catalogs. This was not the case for the other bands. This reinforces the idea that a first-order approximation might not mitigate biases within requirements consistently across redshift for the wide filter. Additionally, an average redshift-bin correction completely fails to stay within the relaxed requirement. Even though the additive biases seem to be well-mitigated in comparison, the results for the multiplicative biases further highlight the challenge in mitigation of chromatic biases for the wide filter, even with full knowledge of the SEDs and PSF model.

\subsection{Galaxy Color Gradients}\label{color_gradients}
Although not shown in Fig.~\ref{fig:corr_bias}, we also estimate the shear bias when using the composite galaxy PSF, constructed from the composite galaxy SED. Technically, a single true PSF for a galaxy with a spatially varying SED (e.g., a bulge$+$disk galaxy each with their own SED) is not defined in the case of a wavelength-dependent PSF \citep{Voigt_2012}. Therefore, using the composite galaxy PSF as an approximation allows us to test the impact of galaxy color gradients in the context of chromatic PSFs. We find that the bias varies between \textit{Diffsky} and cosmoDC2 galaxies, as well as by filter and redshift.

We report upper limits based on the worst case across all redshift bins in Table.~\ref{tab: color_grad}. Note that there is an implicit assumption in these results, which is that AnaCal is an unbiased shear estimator at the level we are testing it \citep{Li_2023}, so any bias here is due to color gradients. We note that in all cases for the WL bands, the bias lies below the strictest survey requirements. The bias for the wide filter varies considerably when we compared \texttt{Diffsky} and cosmoDC2, with the upper limit from cosmoDC2 exceeding the strictest SRD requirements and approaching the relaxed requirement. This once again highlights the sensitivity of the wide filter to chromatic effects. We conclude from these results that galaxy color gradients are unlikely to be an issue for the \textit{Roman} WL bands, but might require additional attention in the case of the wide filter. 

\begin{table}
\begin{adjustbox}{width=0.7\columnwidth,center}
\begin{tabular}{ c|c|c| }

\multicolumn{3}{c}{Upper limit on $10^4|m|$ from} \\
\multicolumn{3}{c}{Galaxy Color Gradients} \\
\hline
Filter &  \texttt{Diffsky} & cosmoDC2\\
\hline
\hline

Y106 &  $0.7$  &$1.4$  \\

J129  & $0.7$  &  $1$   \\

H158  &  $0.7$ & $0.6$  \\

F184 &   $0.6$ & $0.7$  \\

W146 &  $1.5$ & $6.5$  \\

\hline
\end{tabular}
\end{adjustbox}
\caption {The upper limit on the scaled absolute value of the multiplicative bias, $10^{4} |m|$, from galaxy color gradient for both \texttt{Diffsky} and cosmoDC2. The upper limits are set using the worst case value across all redshift bins for each filter. Values for the WL bands $(Y, J, H, F)$ are within the \textit{Roman} SRD requirement of $|m| < 3.2\times 10^{-4}$, while the wide filter exceeds this limit in the case of cosmoDC2. 
}
\label{tab: color_grad}
\end{table}

\section{Realistic Implementation}\label{Results}

We have shown in Sec.~\ref{Mitigation} that a perfect first-order correction to the PSF biases can mitigate chromatic biases within survey requirements for all 4 WL bands, and most likely requires higher order terms for the wide filter W146. This used the full SED information from both stars and galaxies, which we will not have for \textit{Roman's} HLWAS. Therefore, a realistic implementation of the methods described here will rely on developing ways to estimate the SED slopes for galaxies and stars for the \textit{Roman} WL bands using available photometric data. This may be a challenge, as we cannot directly measure the SED inside of any one filter using \textit{Roman} imaging data alone. Therefore, we will rely on obtaining approximate information on what is happening inside each filter by using the photometric information from other imaging bands, along with  assumptions about how the SEDs vary across filters.

The most obvious piece of information about SED slopes from imaging surveys is the measured colors of objects. The color could be treated as a first-order approximation to the SED wavelength dependence across two different filters, meaning it can be used to estimate the SED slope within a filter given some simplifying assumptions. The observed galaxy colors can therefore be used to estimate or learn the chromatic correction coefficients, $\Delta S_1$, defined in  Sec.~\ref{Mitigation}. Here we describe two estimation methods, one analytic and one ML-based using self-organizing maps \citep[SOMs;][]{Kohonen_1982}, that use the photometric information from \textit{Roman} and LSST to estimate these coefficients. We look to predict these coefficients only for galaxies, and assume the true coefficients for stars are known. This is because we expect to have a much better understanding of stellar SEDs, both theoretically and empirically through available spectra. Therefore, we look to predict only the SED slopes of galaxies. We evaluate the performance of both methods on shear measurement for \textit{Roman's} H158 filter as an example case. Finally, we discuss and test the impact of the SED library on the results.

\subsection{Analytical method}
We provide a method to estimate the SED slopes using imaging data directly by assuming the SED can be approximated as linear across adjacent filters -- meaning that the SED slope describing the linear relationship is the same across adjacent filters. We then estimate the SED slope using the flux information from the filter of interest and any adjacent filters. This can technically be done with any two (or more filters), even if they are not adjacent, but the assumption of linearity becomes increasingly inaccurate as filters  with large wavelength separations are used. If we model the SED through some filter $f$ as previously done: 
\begin{equation}\label{SED_linear}
  \text{SED}_f (\lambda) = m_{f} \left( \lambda -  \lambda_0^f\right) + b_{f}   
\end{equation}
where $\lambda_0^f$ is the effective wavelength of filter $f$, and $m_{f}$ and $b_{f}$ are the slope and y-intercept of the SED through the filter. We assume the SED slope is the same for the adjacent filter, $t$. The SED slope can then be estimated as:
\begin{equation}\label{SED_slope}
  \hat{m}_f = \frac{\text{SED} \left(\lambda_0^t\right) - \text{SED} \left(\lambda_0^f\right) }{\lambda_0^t - \lambda_0^f} .  
\end{equation}
Since we do not have access to $\text{SED} \left(\lambda_0^f \right )$ and $\text{SED}\left(\lambda_0^t \right  )$ directly, we estimate them from the flux through the filter $f$, $N_f$, via:
\begin{align}
  N_f &=\int \mathrm{d} \lambda \; F_f(\lambda)\;\text{SED} (\lambda)  \\
  &=  m_f \int \mathrm{d} \lambda \; F_f(\lambda) \; (\lambda - \lambda_0^f) + b_f \int \mathrm{d} \lambda \;  F_f(\lambda)\nonumber
\end{align}
where $F_f(\lambda)$ is the filter throughput of filter $f$. 
We show in Appendix~\ref{appendix:linear_proof}  
that according to the definition of effective wavelength, $\int \mathrm{d}\lambda \; F_f(\lambda) \; \left(\lambda - \lambda_0^f\right )= 0$, simplifying the above equation to:
\begin{equation}
     N_f = b_f \int \mathrm{d} \lambda \;  F_f(\lambda).\\
\end{equation}
Since we can measure $N_f$, the flux in the $f$ band, we can infer $b_f$ via:
\begin{equation}
     b_f = \frac{N_f}{T^f_{0}} \; \text{where } T^f_{n} \equiv\int \mathrm{d} \lambda \;  F_f( \lambda) \;  \left(\lambda - \lambda_0^f\right)^n.
\end{equation}
Here $T_0^f$ can be measured from the filter response function, so a flux measurement $N_f$ uniquely determines the intercept $b_f$ given our assumptions.

We can then estimate the SED value at the effective wavelength of both filters by plugging back into equation~~\eqref{SED_linear}:
\begin{equation}
     \text{SED} \left(\lambda_0^f \right) = \frac{N_f}{T^f_0}, \; \; \text{SED} \left(\lambda_0^t \right  ) =\frac{N_t}{T^t_0}.\\
\end{equation}
Finally, we want to estimate the flux-normalized SED slope, $S^{(1)}$, described in Sec.~\ref{Mitigation:chrom_diff}. To do this, we divide equation~~\eqref{SED_slope} by $N_f$ and plug in the above results to obtain the final analytical estimator:
\begin{equation}\label{anal_estimator}
     \hat{S}^{(1)} \equiv \frac{\hat{m}_f}{N_f} = \frac{1}{\lambda_0^t - \lambda_0^f} \frac{1}{N_f} \frac{T^f_0N_t - T^t_0 N_f }{ T^t_0 T^f_0 }.
\end{equation}
As we can see, this estimator depends solely on the measured flux and the transmission curves for each filter. In this context, the flux $N$ refers to the observed flux rather than the true flux. 
We note that the term $N_t/N_f$ resembles something like the color, which is proportional to $\log_{10}(N_t/N_f)$. Therefore, this formalism contains the same information as the color. For low SNR objects this ratio might diverge, leading to extreme over correction. In such cases, an alternative might be to assign those objects a chosen average value (either from the total sample or their assigned redshift bin).

\begin{figure}
  \center

  \includegraphics[width=0.45\textwidth]{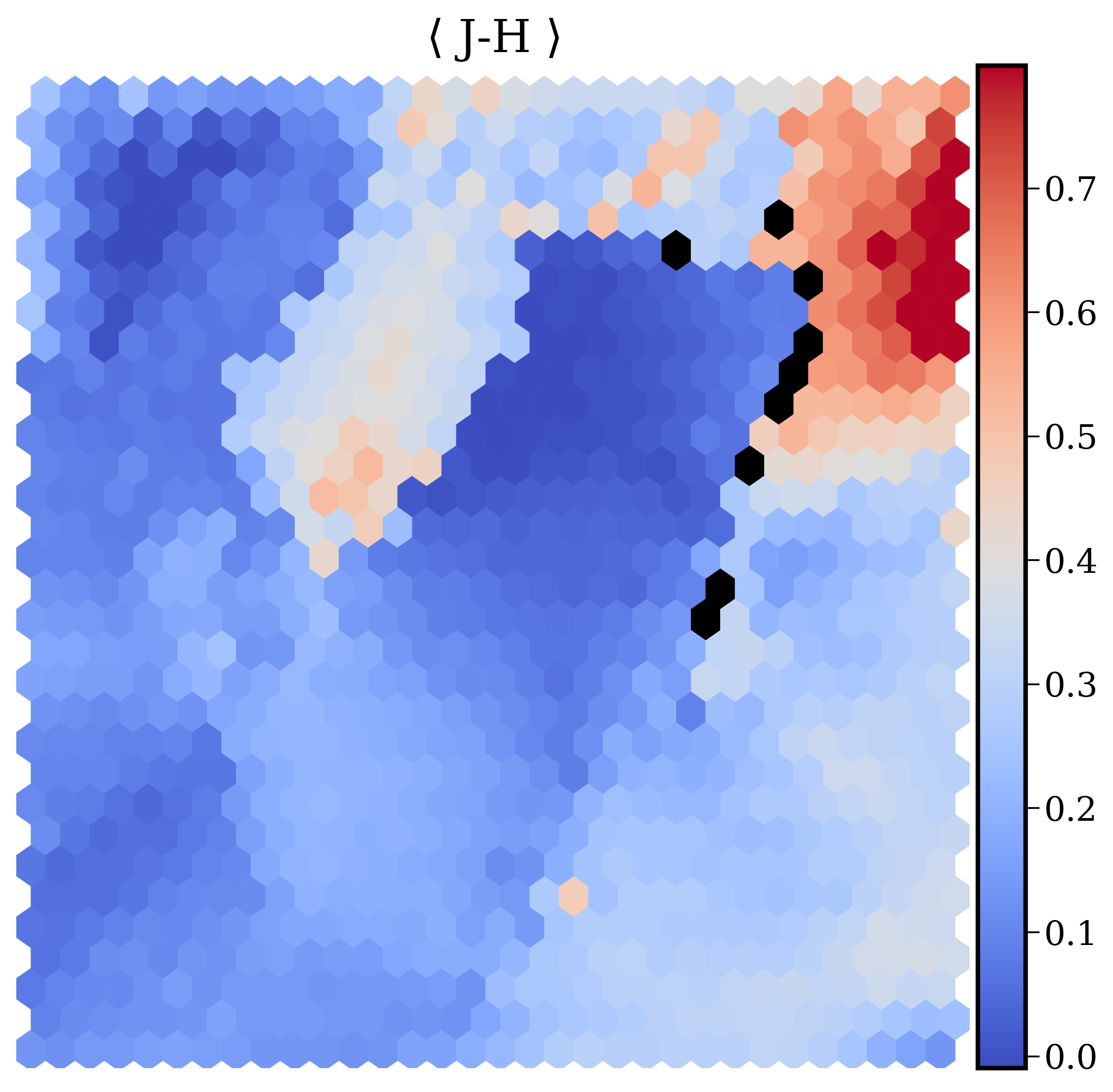}
  \includegraphics[width=0.45\textwidth]{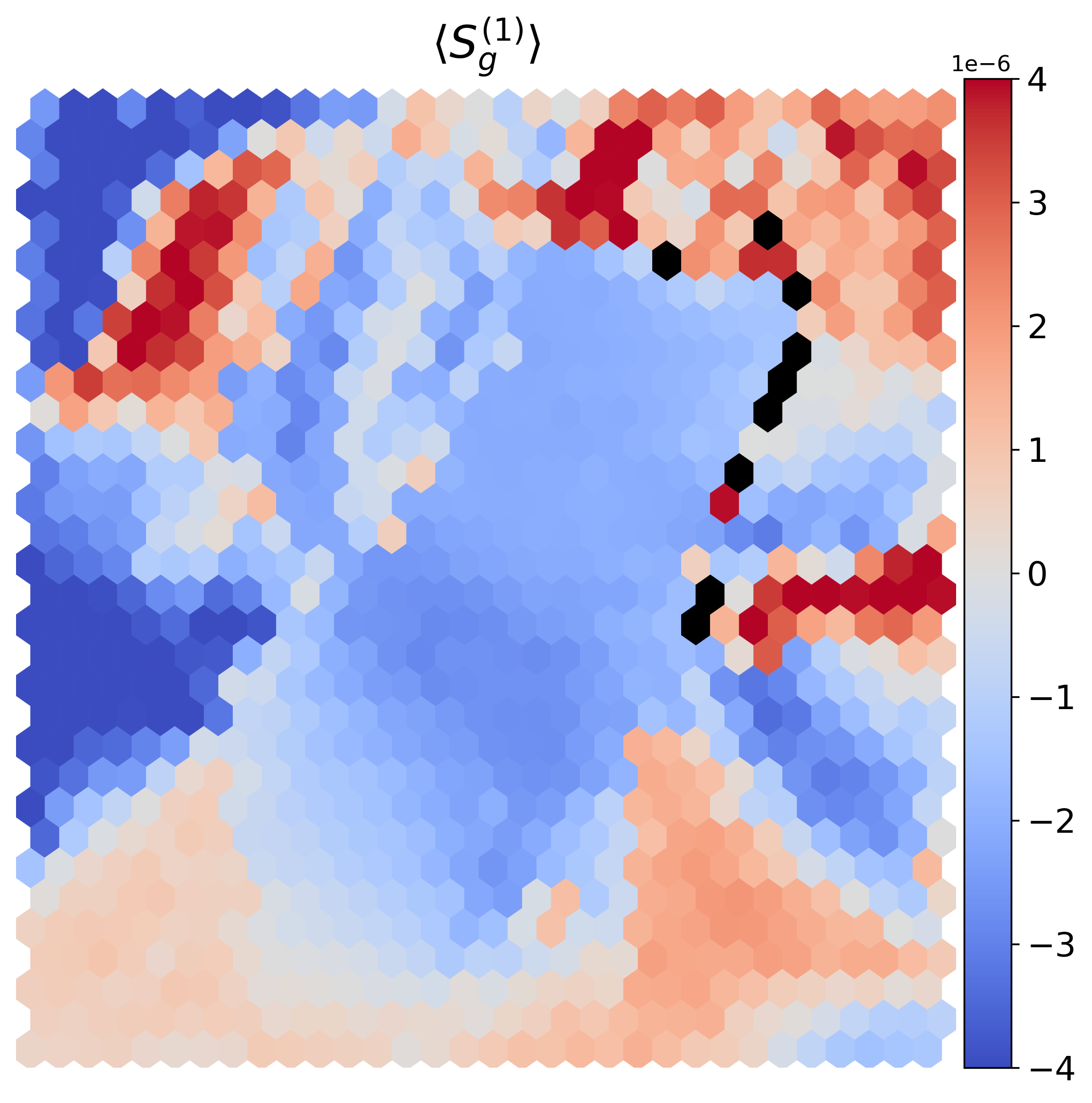}
  \caption{Trained SOM for the \texttt{Diffsky} galaxies using 9 colors constructed from the LSST (\textit{ugrizy}) + \textit{Roman} (\textit{YJHF}) bands. The top and bottom panels show the $\langle J-H \rangle $ color and $\langle S_g^{1} \rangle$ values for the training set galaxies, respectively. 
  The color distribution is somewhat smooth across the plane, with some visible discontinuities for redder galaxies. The distribution of $S_g^{1}$, the galaxy SED slope, also exhibits discontinuous regions, highlighting that a single color can fail to predict the SED behavior within the filter. 
  }
  \label{fig: SOMs}

\end{figure}

\begin{figure*}
    \includegraphics[width=0.95\linewidth]{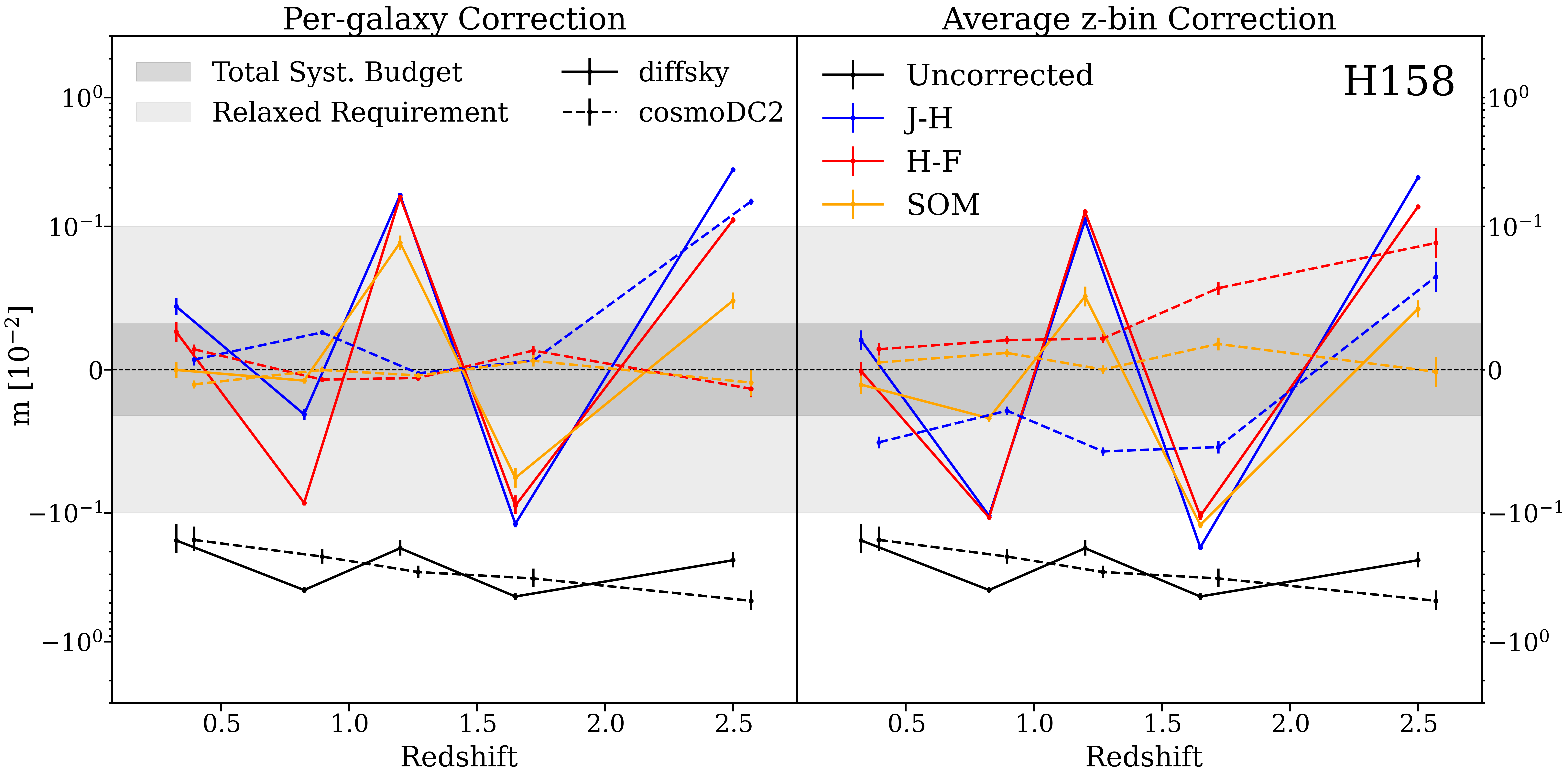}

    \caption{The multiplicative shear bias for filter H158 after the PSF-level correction using the estimated values of $\Delta\hat{S}_1$ for both \texttt{Diffsky} (solid) and cosmoDC2 (dashed). $\Delta\hat{S}_1$ is estimated in two ways: analytically using imaging data from adjacent filters J129 (blue) and F184 (red), and through dimensionality reduction of the LSST$+$Roman colors using a SOM (orange). The left panel shows the results after the per-galaxy correction, where each galaxy is corrected using its individual estimate of $\Delta\hat{S}_1$. The right panel shows the results after applying an average correction, where each galaxy's PSF is corrected using the average coefficient, $\langle\Delta\hat{S}_1\rangle_z$, for its redshift bin. The results with \texttt{Diffsky} reveal biases that often exceed the requirements when using the analytical methods, while the SOM performs much better -- with the exception of the redshift bin $1.4<z<1.9$ in the average correction scheme. Results for cosmoDC2 are more optimistic when using either estimator, consistently staying within the most stringent requirement in the per-galaxy case (with the exception of $J-H$ for the highest redshift bin) and within the relaxed requirement in the average case.
    }

    \label{fig:real_corr}
\end{figure*}

\begin{table}
\begin{adjustbox}{width=1.0\columnwidth,center}
\begin{tabular}{ |c|c|c|c|c| } 

\multicolumn{1}{|c|}{} & \multicolumn{4}{c|}{\large{Mean Relative Error}} \\

\multicolumn{5}{|c|}{} \\

\multicolumn{1}{|c|}{\normalsize{\textbf{\texttt{Diffsky}}}} \\
\hline

 &  J-H & H-F & SOM & Allowed for $|m| < 0.001$\\
\hline
\hline

All galaxies &  20\%  & 18\%  &  8\% & 41\%\\
\hline
$z < 0.65$   & 60\%  &  42\% & 33\% & 61\%   \\

$0.65< z < 1$   &  -22\% &  -23\% & -3\% & 25\%  \\

$1< z < 1.4$   &  \textbf{\textcolor{red}{187\%}}  &  \textbf{\textcolor{red}{171\%}} & \textbf{\textcolor{red}{187\%}} & 53\%  \\
$1.4< z < 1.9$   &  \textbf{\textcolor{red}{-39\%}}  & -16\% & 21\% & 22\%  \\

$1.9< z < 3.1$   &  \textbf{\textcolor{red}{107\%}}  & \textbf{\textcolor{red}{64\%}} & 21\% & 42\%  \\

\multicolumn{5}{|c|}{}\\
\multicolumn{1}{|c|}{\textbf{\normalsize{cosmoDC2}}} \\
    \hline
 &  J-H & H-F & SOM & Allowed for $|m| < 0.001$\\
\hline
\hline

All galaxies &  -2\%  &   26\% &  16\% & 45\%\\
\hline

$z < 0.65$   & -5\%  &  50\% & 41\% & 62\%   \\

$0.65< z < 1$   &  -1\%  &  26\% & 21\% & 46\%  \\

$1< z < 1.4$   &   -10\%  &  20\% & 12\% & 35\%  \\
$1.4< z < 1.9$   &  -7\%  &  30\% & 17\% & 31\%  \\

$1.9< z < 3.1$   & 11\%  &  16\% & -3\% & 21\%  \\

\hline
\end{tabular}
\end{adjustbox}
\caption {The mean relative error, defined in equation~~\eqref{eq:mean_rel_err}, for each estimator of $\Delta\hat{S}_1$. Values are provided for the entire test sample (all galaxies) and each redshift bin. Results from both \texttt{Diffsky} (top section) and cosmoDC2 (bottom section) are shown. The last column shows the ``allowed'' mean relative error for a multiplicative bias requirement of $|m| < 0.001$. This metric shows an approximate margin of error for the mean of the correction coefficient if we wish to stay within a relaxed requirement. Values in red are those that exceed this allowed error margin and therefore might exceed this requirement. We see systematically higher residual biases for correction with \texttt{Diffsky} than cosmoDC2, with multiple redshift bins failing to stay within the predicted allowed accuracy limits. Results with cosmoDC2 consistently stay within the limits, with a few cases living close to the allowed error margin. }
\label{tab:performance}
\end{table}

\subsection{Self-organizing maps}

SOMs are a type of unsupervised learning where high-dimensional data is mapped onto a lower-dimensional (typically 2D) grid while preserving its topological structure. SOMs have been widely applied to dimensionality reduction problems. More recently they have been popularly applied to reduce  the high-dimensional galaxy color space for redshift estimation \citep[e.g,][]{Myles_2021, Campos_2024}. In this work we have a very similar problem, where the multi-band observed colors are correlated with the SED slopes we are interested in.

We use the SOM implementation in \texttt{RAIL}\footnote{\url{https://github.com/LSSTDESC/RAIL}} to reduce the 9-dimensional color space $(u-g, g-r, r-i, i-z, z-y, y-Y, Y-J, J-H, H-F)$, from the catalog measurements of the LSST$+$\textit{Roman} bands, into a 2D grid. To do this, we need both a training and testing set. The testing set consists of the same 10,000 galaxies previously used for image simulations and tests. The training set consists of a sample of 40,000 galaxies, distinct from those used in testing, from the same extragalactic catalog used for the test set. The observed magnitudes of these galaxies are calculated assuming they come from a deep-drilling field. This is because we expect to have spectra from these fields, from which SED slopes can be measured. Therefore, using magnitudes from a deep field makes more sense for the training sample. In terms of the catalog-level noise generated to produce these deeper observed magnitudes, we only change the 5$\sigma$ limiting magnitudes used for the \texttt{PhotErr} error model described in Sec.~\ref{Simulations:noise}. The LSST deep field 5$\sigma$ point limiting magnitudes are taken from \cite{Gris_2024}, while the \textit{Roman} ones come from the WFI technical page\footnote{\url{https://roman.gsfc.nasa.gov/science/WFI_technical.html}} (values used are for 1 hour integration time).

We now detail the parameters that go into the SOM. We follow the example provided in the \texttt{RAIL} documentation\footnote{ \url{https://rail-hub.readthedocs.io/projects/rail-notebooks/en/latest/rendered/estimation_examples/somocluSOM_demo.html}} to initialize the SOM. We use a hexagonal grid type with \texttt{std\_coeff = 12.0} and \texttt{som\_learning\_rate=0.75}. The only change we make is to reduce the number of rows and columns to 32 each, since using the value provided in the example produced a lot of empty cells. The SOM is then trained using the 9 colors from LSST and \textit{Roman} photometry. Each cell is then assigned the average value of $S^{(1)}_g$ for the galaxies in that cell. The galaxies in the test set are then assigned the average value of the cell they fall into to estimate $\hat{S}^{(1)}_g$. If galaxies fall into unpopulated SOM cells they are assigned the average training set value for $S^{(1)}_g$. 
This only occurred for \texttt{Diffsky}, with 18 out of 10,000 galaxies falling in unpopulated cells. 

Fig.~\ref{fig: SOMs} shows the trained SOM for \texttt{Diffsky} only, for visualization. We see that the color distribution in the top panel is fairly continuous, with a few clear discontinuities for redder galaxies. The distribution of values for $\langle S_g^{(1)}\rangle$ 
can be more discontinuous, hinting that individual colors may not always smoothly predict the SED behavior within the filter.

\subsection{Performance}

We evaluate the performance of the analytical and SOM methods for mitigating chromatic PSF biases. We calculate the chromatic bias coefficients $\Delta\hat{S}_1$ using the estimated galaxy SED slopes, $\hat{S}_{g}^{(1)}$, and the true stellar SED slopes $S_{\star}^{(1)}$. For the analytical estimator, we compute the values of $\Delta\hat{S}_1$ using the two adjacent filters to H158: J129 and F184. The corrections using J129 and F184 are denoted as $J-H$ and $H-F$,  
respectively. We begin by evaluating the accuracy of each estimator by calculating the mean relative error:
\begin{equation}\label{eq:mean_rel_err}
    \text{Mean Relative Error} =  \frac{\langle \Delta\hat{S}_1 - \Delta S_1 \rangle}{\langle\Delta S_1\rangle}.
\end{equation}
Recall that $\Delta\hat{S}_1$ is calculated from the noisy magnitudes while $\Delta{S}_1$ is calculated from the true SEDs (no noise). 
Both make a linearity assumption about the galaxy SED: $\Delta\hat{S}_1$ for the analytical method assumes linearity across the filter pair, while $\Delta\hat{S}_1$ for the SOM method and $\Delta{S}_1$ assume linearity only within the reference filter. Values for each estimator are provided in Table~\ref{tab:performance} for the entire test sample and for each redshift bin. Results for \texttt{Diffsky} and cosmoDC2 are shown in the top and bottom sections of the table, respectively. The last column shows the ``allowed'' mean relative error for a requirement of $|m| < 0.001$. This value is calculated by taking the ratio $0.001/|m|$, where $m$ is the measured multiplicative bias for each row from the results in Sec.~\ref{Bias}. 
This shows an approximate margin of error for the mean of the correction coefficient if we wish to stay within a relaxed requirement. Values in red are those that exceed this allowed error margin and therefore might fall out of this requirement. Note that the distribution of $\Delta S_1$ is not Gaussian for every subsample, and therefore the mean relative error might not be directly proportional to the bias.

\begin{figure}
  \center

  \includegraphics[width=0.45\textwidth]{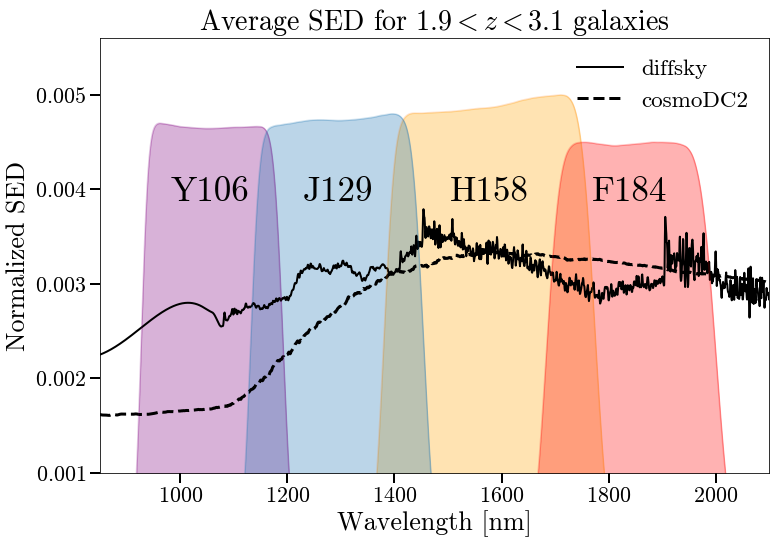}
  \caption{The average flux-normalized galaxy SED for \texttt{Diffsky} (solid line) and cosmoDC2 (dashed line) for the highest redshift bin. The filled areas show the filter transmission curves of the 4 WL filters for visualization. We see a large difference between the average SEDs from both catalogs, with the average cosmoDC2 SED being closer to linear both within and across filters compared to the  \texttt{Diffsky} SEDs.}
  \label{fig: avg_sed}

\end{figure}

From the table we see that the results for both \texttt{Diffsky} and cosmoDC2 can stay within limits when averaging over all galaxies using either correction method. \texttt{Diffsky} has multiple redshift bins, however, that fail to exceed the allowed relative error after correction, with redshift bin $1<z<1.4$ being the worst case regardless of the method, and bin $1.9<z<3.1$ failing for the analytical methods. On the other hand, the methods work much better for cosmoDC2, never exceeding the approximate limit. This difference in performance between catalogs is due to their intrinsic differences in SED libraries. SEDs for the galaxies in our magnitude range in the cosmoDC2 catalog are more linear (on average) within the \textit{Roman} filters than those in \texttt{Diffsky}. This difference is shown in Fig.~\ref{fig: avg_sed}, where we show the average SED for both catalogs for the highest redshift bin. As we can see, there is a big difference between the linearity assumption across adjacent filters when comparing \texttt{Diffsky} and cosmoDC2. This means that predicting the SED slopes inside a filter, either analytically or through ML, using imaging data becomes less accurate. This makes sense: if the SED is doing something completely different in adjacent filters, the assumptions going into the analytical method no longer hold. Using a SOM improves the overall accuracy in this case for \texttt{Diffsky}, but is still not perfect. We now test the accuracy on image simulations to get a better idea of the direct performance in shear measurement.

\begin{figure*}
    \includegraphics[width=0.7\linewidth]{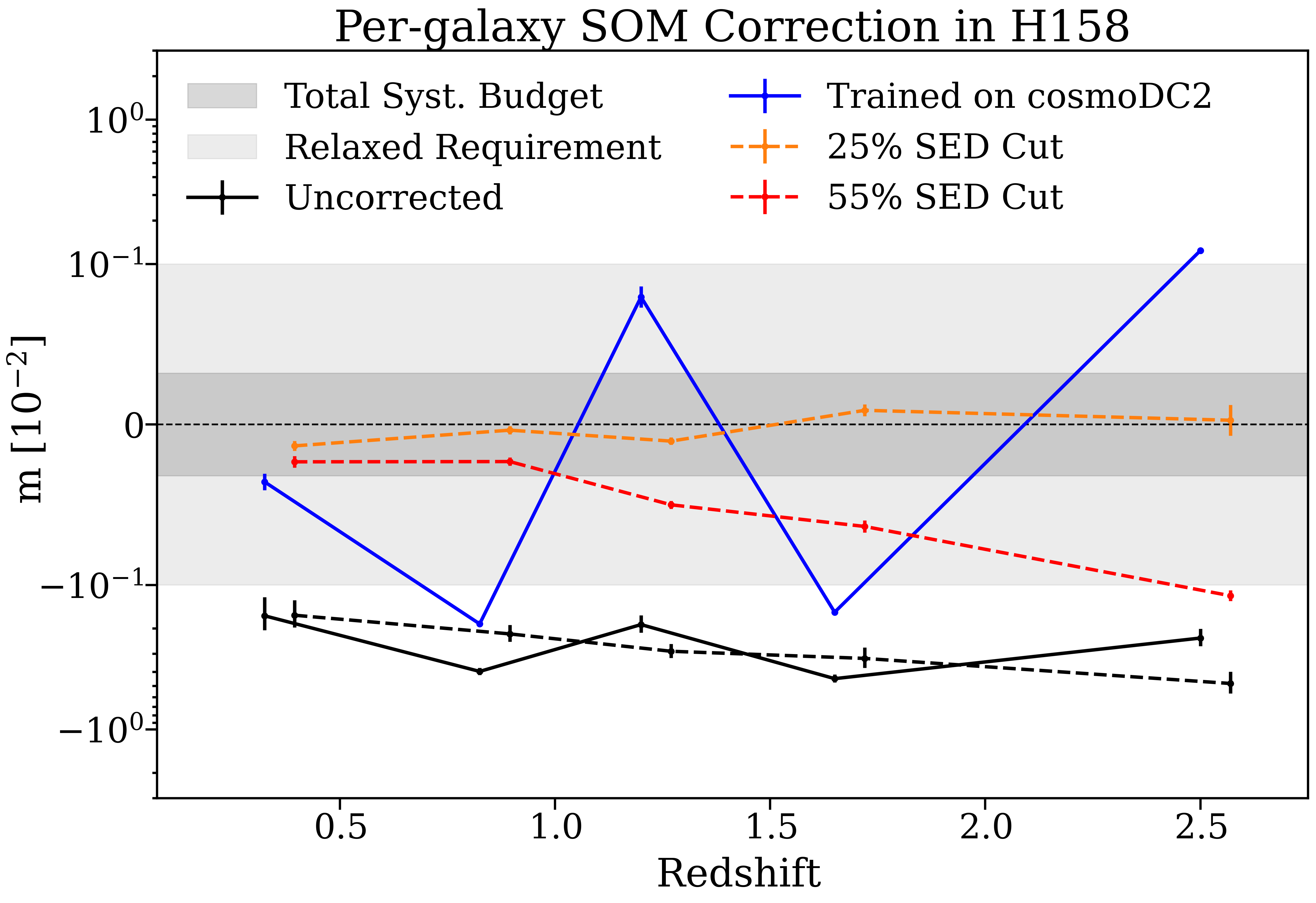}

    \caption{The multiplicative shear bias for two tests of the impact of the SED library on the chromatic correction using SOMs to learn the SED-dependent correction coefficients. The two scenarios are: train the SOM on cosmoDC2 and test on \texttt{Diffsky} (blue solid line), and two different SED removal cuts: 25\% (orange dashed line) and 55\% (red dashed line). As with previous figures, solid (dashed) lines are used when the test set of image simulations comes from \texttt{Diffsky} (cosmoDC2). 
    The SED removal cuts involve systematically removing certain SED templates from the training sample until $25\%/55\%$ of galaxies in the test sample are no longer represented in the training sample. Solid and dashed lines represent \texttt{Diffsky} and cosmoDC2 test set galaxies, respectively. As we can see, training on cosmoDC2 and testing on \texttt{Diffsky} results in a decrease in performance compared to training with and testing on \texttt{Diffsky}. The correction lies outside the relaxed requirement for most redshift bins, highlighting the sensitivity of the SOM method to the training set. In the case of the 25\% SED removal cut, we see no effect as the correction is nearly perfect. Even in the case of of removing 55\%, we see a failure to stay within the relaxed requirement only for the highest redshift bin. The results give some understanding of the impact the SED library can have on the quality of a SOM-based chromatic PSF correction. }

    \label{fig:real_corr_SOMSEDImpact}
\end{figure*}

Figure.~\ref{fig:real_corr} shows the multiplicative shear bias using the estimated $\Delta\hat{S}_1$ values for both \texttt{Diffsky} and cosmoDC2. We show both analytical estimators, $J-H$  and $H-F$, as well as the estimator using SOMs. The left panel shows the results after applying a per-galaxy correction, where each galaxy is corrected using its individual estimate of $\Delta\hat{S}_1$. The right panel shows the average correction, where each galaxy's PSF is corrected using the average coefficient, $\langle\Delta\hat{S}_1\rangle_z$, for its particular redshift bin.  The results with \texttt{Diffsky} reveal biases that often exceed the requirements when using the analytical methods, while the SOM performs much better -- with the exception of the redshift bin $1.4<z<1.9$ in the average correction scheme.

For the cosmoDC2 per-galaxy correction, all methods stay within requirements except for the analytical correction using $J-H$ for the highest redshift bin. We explain this difference between the $J-H$ and $H-F$ estimators by looking at Fig.~\ref{fig: avg_sed}, where we clearly see that the average SED slope is relatively similar between H158 and F184, but consistently different between J129 and H158 for cosmoDC2. 
For the average correction schemes, cosmoDC2 never lies outside the relaxed requirement, but is dangerously close to the limit for $J-H$ and $H-F$ in the highest redshift bin. The SOM method vastly improves the correction and stays even within the most stringent requirement. However, we note that the improved performance of the SOM method also depends on the fact that the SED library used in the training and test set is the same. Here the training sample perfectly represents the testing sample, but in real data we are unlikely to have perfect completeness in SED space. 

The results here show the performance of two empirical methods of estimating the SED slope. For the case of \texttt{Diffsky}, where galaxy SEDs are highly non-linear, the analytical method presented here does not mitigate biases consistently across redshift. The SOM, which make weaker assumption about galaxy SEDs, proved more effective at estimating chromatic PSF corrections and mitigating shear biases. However, results using SOMs are highly dependent on having a representative SED training sample, which might not be the case with real data. For the case of cosmoDC2, we see that both the analytic and SOM methods mitigate biases within a relaxed requirement across redshift, with a single exception for high redshift galaxies using $J-H$. However, the success of the analytical estimator highly depends on the assumptions about galaxy SEDs holding for this particular catalog, which might not be the case with real galaxies.

\subsection{Impact of SED library}
We observe that the performance of our chromatic bias correction depends on the choice of the SED library, motivating further investigation into its impact. To assess this, we conduct two tests: First, we train the self-organizing map (SOM) using the cosmoDC2 galaxies while testing on the \texttt{Diffsky} sample, introducing mismatches between the training and testing set.  Second, we remove a selected fraction of the SED templates from the cosmoDC2 training sample while keeping them in the test sample, introducing incompleteness in the training set. 
Since \texttt{Diffsky} lacks predefined SED groupings, the second test is performed only on cosmoDC2.

To systematically remove SED templates, we first apply a K-means clustering algorithm to identify groups of similar SEDs within the cosmoDC2 library. We divide the SEDs into ten clusters and selectively remove entire clusters to control the fraction of missing templates and to ensure that templates are removed systematically rather than at random. Two removal cases are considered: one eliminating 25\% (98 templates) of galaxies and another removing 55\% (159 templates). 

Figure~\ref{fig:real_corr_SOMSEDImpact} illustrates the impact of these choices. Training on cosmoDC2 and testing on \texttt{Diffsky} (blue solid line) results in a noticeable degradation in correction performance compared to training and testing on the same dataset, with corrections failing to meet the relaxed survey requirements in most redshift bins. For the SED removal tests, we observe no significant effect when 25\% of SEDs are removed. However, when 55\% of the templates are excluded, a substantial performance drop occurs at the highest redshifts. We verify that this degradation is not simply due to removing a disproportionate number of high-redshift galaxies from the training set. These findings highlight the sensitivity of SOM-based corrections to the representativeness of the training SED population. If the training sample is significantly different from the observed galaxy population, correction performance can vary unpredictably. Conversely, when the training set is merely incomplete but still representative, the impact appears to be less severe.  Fortunately, in reality we expect to have some spectroscopy for the \textit{Roman} galaxies that can be used to build a training sample that is likely incomplete but at least represents some fraction of the sample accurately.

\section{Conclusion}\label{Conclusion}


In this work, we investigated the impact of chromatic biases introduced by the SED differences between stars, which are used to model the PSF, and galaxies, which are used for shear measurements, in the context of the \textit{Roman} weak lensing survey. We quantified the magnitude of these biases using realistic \textit{Roman}-like image simulations and developed mitigation strategies aimed at reducing systematic biases in weak lensing shear inference to within mission requirements.

Our analysis demonstrated that multiplicative biases in the ensemble shear estimation due to chromatic PSF effects in the weak lensing bands (Y106, J129, H158, and F184) reach approximately 0.2\%, while biases in the wide filter (W146) can reach 2\%. These exceed the mission requirement of $|m| < 0.032\%$ and the relaxed requirement of $|m| < 0.1\%$, highlighting the necessity of robust corrections. We note that the bias for some redshift bins can reach 0.4-0.9\% for the WL bands and 3-6\% for the wide filter. Additive biases remain within acceptable levels in the weak lensing bands but exceed the systematic error budget in the wide filter, further underscoring the challenges posed by chromatic biases in a potential shear analysis with the wide filter.

To address these biases, we implemented and tested PSF-level correction methods leveraging the well-characterized nature of the \textit{Roman} PSF. We demonstrated that a first-order correction, when applied under idealized conditions with perfect SED knowledge, is capable of mitigating biases to within survey requirements for the weak lensing bands. However, for the wide filter, the complexity of chromatic variations requires higher-order corrections, as residual biases remain significant even after applying first-order corrections. This once again shows the challenges the wide filter poses for WL, even with perfect knowledge of galaxy SEDs. We also quantify the impact of galaxy color gradients using the composite galaxy SED to construct an approximation of the galaxy PSF. We find a multiplicative bias that does not exceed $10^{-4}$ across all redshift bins for the WL bands (with the exception of the $Y$ band for cosmoDC2 galaxies). The biases are larger for the wide filter and can exceed the \textit{Roman} SRD requirements. We conclude that galaxy color gradients are unlikely to be a concern for the \textit{Roman} WL bands, but may need further consideration for the wide filter.

To make the method applicable to real data, we explored two approaches for empirically estimating the necessary SED-based corrections: analytical color-based estimators and machine learning techniques using SOMs. We found that both methods effectively reduce biases, but their success is strongly dependent on the choice of SED library. In particular, the performance of the SOM approach is sensitive to the representativeness of the training set, with significant degradation in cases where the SED distribution of the test sample deviates from that of the training sample. These findings indicate that future weak lensing surveys must carefully consider the choice of SED priors when designing chromatic bias mitigation strategies. The analytical approach provides a fully empirical way of determining the correction coefficients needed for our mitigation method. Although reliant on certain assumptions about the galaxy SEDs, this method proved robust even in the case of semi-realistic noisy magnitudes.

Future work should focus on refining mitigation methods in more realistic observational conditions. This includes incorporating measurement noise, calibration using spectroscopic data from the \textit{Roman} deep tier, imperfect knowledge of galaxy redshifts, impact of image coaddition, all of which could impact the accuracy of the chromatic bias corrections. Additionally, optimizing for different survey scenarios will be needed for such a correction. Given that \textit{Roman's} survey strategy is not finalized, understanding the implications of different survey scenarios on the quality of the chromatic correction will be important. We emphasize that this work has produced results for oversampled images. Therefore, it will be extremely important to understand how the image coaddition of individual undersampled exposures will affect the chromaticity of the coadded PSF. In terms of practical the implementation of our work, while the correction presented here is applied on a per-galaxy basis, it can also be modified to an ensemble-level correction and extended to work for \texttt{METACALIBRATION} \citep{Huff_2017, Sheldon_2017} and \texttt{METADETECTION} \citep{Sheldon_2020}. Finally, understanding the range of training data used for SOM-based approaches, particularly with spectroscopic data, may improve generalization to real survey conditions.

In summary, this work provides a systematic analysis of chromatic biases in weak lensing shear measurements for \textit{Roman} and presents a framework for mitigating these effects. By quantifying biases, testing mitigation strategies, and outlining their limitations, we establish a foundation for future weak lensing analyses that require high-precision PSF modeling. As the field moves toward increasingly stringent systematics control, continued efforts in modeling and correcting chromatic PSF effects will be vital to ensuring the scientific success of next-generation weak lensing surveys.


\section*{Acknowledgements}

This paper has undergone internal review in the \textit{Roman} High Latitude Imaging Survey Cosmology Project Infrastructure Team (PIT). We would like to thank Brett Andrews, Kaili Cao, Katherine Laliotis, Christopher Hirata, Arun Kannawadi, and Ami Choi for helpful comments and feedback during the review process. We also thank Andrew Hearin for reviewing our presentation of the galaxy model and how the results depend on the choice of mock.

This work was supported in part by the OpenUniverse effort, which is funded by NASA under JPL Contract Task 70-711320, “Maximizing Science Exploitation of Simulated Cosmological Survey Data Across Surveys”; and in part by the “Maximizing Cosmological Science with the Roman High Latitude Imaging Survey” Roman Project Infrastructure Team (NASA grant 22-ROMAN11-0011).
Xiangchong Li is an employee of Brookhaven Science Associates, LLC under
Contract No. DE-SC0012704 with the U.S. Department of Energy.
TZ is supported by Schmidt Sciences.

\section*{Data Availability}
The \texttt{Diffsky} extragalactic catalog and stellar catalog are available through the NASA/IPAC
Infrared Science Archive (IRSA) at \url{https://irsa.ipac.caltech.edu/data/theory/openuniverse2024/overview.html}. The cosmoDC2 catalog is available  through NERSC at \url{https://portal.nersc.gov/project/lsst/cosmoDC2/_README.html}. The \texttt{GalSim} package is publicly available at \url{https://github.com/GalSim-developers/GalSim}. The \texttt{PhotErr} package is publicly available at \url{https://github.com/jfcrenshaw/photerr}. The code used to generate the image simulations and analysis for this work is available at \url{https://github.com/FedericoBerlfein/RomanChromaticPSF/tree/main}.



\bibliographystyle{mnras}
\bibliography{main} 




\appendix

\section{Magnitude Errors}\label{appendix:mag_err}
The photometric error model for point sources in \texttt{PhotErr} is based on the work of \cite{Ivezic_2019}, which provides a way of estimating magnitude errors:
\begin{equation}\label{eq: point_magerr}
\sigma^2_{m, \text{point}} = (0.04 - \gamma)x + \gamma x^2
\end{equation}
where $\sigma_{m, \text{point}}$ is the magnitude error, $x = 10^{0.4 (m - m_5)}$
where $m$ is the magnitude and $m_5$ the $5\sigma$ limiting magnitude, and the value of $\gamma$ represents the impact of sources of noise (e.g. sky brightness).
$\gamma$ is set to $0.04$ for \textit{Roman} in \cite{Graham_2020} and has a similar value of 0.039 for LSST \citep{Ivezic_2019}.
For extended objects, \texttt{Photerr} follows the methodology described by \cite{van_den_Busch_2020}, which accounts for the increased complexity of measuring fluxes in extended sources assuming a Gaussian PSF and elliptical aperture, and Sersic galaxy profile with disk $+$ bulge components:
\begin{equation}\label{eq: ext_magerr}
\text{SNR}_{\text{ext}} \propto \text{SNR}_{\text{point}} \sqrt\frac{A_{\text{PSF}}}{A_{\text{ap}}}
\end{equation}
where $A_{\text{ap}}$ and $A_{\text{PSF}}$ are the area of the aperture and PSF, respectively. \texttt{PhotErr} sets the proportionality constant to unity to recover the error for a point source when $A_{\text{ap}} \rightarrow A_{\text{PSF}}$. $A_{\text{PSF}}$ is calculated using the input parameters for the PSF FWHM in the error model. $A_{\text{ap}}$ needs information about the galaxy's semi-major ($a_{\text{gal}}$) and semi-minor ($b_{\text{gal}}$) axis to be calculated. Therefore, \texttt{PhotErr} requires the user to provide these two quantities in addition to the true magnitudes and error model described earlier.

The semi-major and semi-minor axis depend on the galaxy half light radius (HLR) and minor-to-major axis ratio, $q = \frac{1-e}{1+e}$, by the following relation:
 \begin{equation}\label{eq: semi_axis}
a_{\text{gal}} = \frac{\text{HLR}}{\sqrt{q}}; \; \; b_{\text{gal}} = \text{HLR}\sqrt{q}.
\end{equation}
 We approximate the semi-major and semi-minor axis for all galaxies using the catalog-level information of each component's (disk/bulge) half light radius (HLR) and ellipticities. The total galaxy HLR and minor-to-major axis ratio are available for cosmoDC2 galaxies, but not for diffsky galaxies, where only the per-component information is available. Therefore, for \texttt{Diffsky}, we estimate an overall $q$ from the component information. We estimate the total galaxy HLR as the flux-average of the disk and bulge HLRs. Similarly, we use the ellipticities ($e_1$ and $e_2$) given in the catalog for the disk and bulge to calculate the magnitude of the total ellipticity, $e = \sqrt{e_1^2 + e_2^2}$, for each component. We then estimate the total galaxy ellipticity as the flux-average of the disk and bulge ellipticities. The total galaxy ellipticity is then used to calculate $q$, and in combination with the galaxy HLR, the semi-major and semi-minor axis. We confirm, using the catalog-level information in cosmoDC2 galaxies for the total HLR and minor-to-major axis ratio, that this approximation does not considerably change the estimates of the semi-major and semi-minor axis or the resulting magnitude errors compared to the case where the catalog values are directly used. 

 \section{Proof of Linear correction}\label{appendix:linear_proof}

For the linear and flux-normalized SED used in Eq.~\eqref{eq:psf_diff_linear}, $S_x = m_x(\lambda - \lambda_0) + b_x$, the flux through the bandpass, $N_x$, is:

\begin{equation}
    N_x = m_x \int d \lambda F(\lambda) (\lambda - \lambda_0) + b_x \int d \lambda F(\lambda).
\end{equation}
Since the SEDs are flux-normalized, $N_x = 1$.
The first term in the equation can be simplified using the definition of the effective wavelength:

\begin{align}
    &\int d \lambda F(\lambda) (\lambda - \lambda_0) =
\int d \lambda F(\lambda) \lambda - \lambda_0 \int d \lambda F(\lambda)\\
&= \int d \lambda F(\lambda) \lambda - \frac{\int d\lambda \; F(\lambda) \lambda}{\int d\lambda \; F(\lambda)} \int d \lambda F(\lambda) = 0
\end{align}
This simplifies the flux to $N_x =  b_x \int d \lambda F(\lambda) = 1$. This means that $b_x = \frac{1}{\int d \lambda F(\lambda)}$ and is therefore constant for all objects. This means that the flux-normalized SED differences in Eq.~\eqref{eq:psf_diff_linear}
only include the first-order term.


\bsp	
\label{lastpage}
\end{document}